\documentclass[preprint,preprintnumbers,amsmath,amssymb,floatfix,nofootinbib]{revtex4}
\usepackage[dvips]{graphicx}
\usepackage{url}
\usepackage{subfigure}
\usepackage{color}

\begin{document}          
\preprint{FSU-HEP-2009-0314}
\preprint{UCLA/09/TEP/49}
%
\title{$W$- and $Z$-boson production with a massive bottom-quark pair at
  the Large Hadron Collider}
\author{F.~Febres~Cordero}
\email{ffebres@physics.ucla.edu}
\affiliation{Department of Physics and Astronomy, UCLA, Los Angeles, CA 90095-1547, USA}
\author{L.~Reina}
\email{reina@hep.fsu.edu}
\affiliation{Physics Department, Florida State University,
Tallahassee, FL 32306-4350, USA}
\author{D.~Wackeroth}
\email{dow@ubpheno.physics.buffalo.edu}
\affiliation{Department of Physics, SUNY at Buffalo,
Buffalo, NY 14260-1500, USA}

\date{\today}

\begin{abstract}
  We present total and differential cross sections for $Wb\bar{b}$ and
  $Zb\bar{b}$ production at the CERN Large Hadron Collider with a
  center-of-mass energy of $\sqrt{s}=14$~TeV, including
  Next-to-Leading Order (NLO) QCD corrections and full bottom-quark
  mass effects.  We also provide numerical results obtained with a
  center-of-mass energy of $\sqrt{s}=10$~TeV. We study the scale
  uncertainty of the total cross sections due to the residual
  renormalization- and factorization-scale dependence of the truncated
  perturbative series.  While in the case of $Zb\bar{b}$ production
  the scale uncertainty of the total cross section is reduced by NLO
  QCD corrections, the $Wb\bar{b}$ production process at NLO in QCD
  still suffers from large scale uncertainties, in particular in the
  \textit{inclusive} case.  We also perform a detailed comparison with a
  calculation that considers massless bottom quarks, as implemented in
  the Monte Carlo program MCFM. The effects of a non-zero bottom-quark
  mass ($m_b$) cannot be neglected in phase-space regions where the
  relevant kinematic observable, such as the transverse momentum of
  the bottom quarks or the invariant mass of the bottom-quark pair,
  are of the order of $m_b$. The effects on the total production cross
  sections are usually smaller than the residual scale uncertainty at
  NLO in QCD.
\end{abstract}
%
\maketitle

\section{Introduction}
\label{sec:intro}
The Large Hadron Collider (LHC) at CERN (Geneva, Switzerland) is
scheduled to start operation by the end of 2009. One of the most
important items on its agenda is the investigation of the mechanism of
electroweak symmetry breaking (EWSB), in particular the discovery of
one or more Higgs bosons. Once discovered, the measurement of the
Higgs bosons' properties will be crucial to unravel the underlying
EWSB mechanism.  The production of a weak gauge boson, $W^\pm$ (from
now on indicated simply as $W$ unless differently specified) or $Z$,
with a pair of bottom ($b$) quarks, contributing to both the $W/Z+1\,b$-jet
and $W/Z+2\,b$-jets signatures, represents both an interesting
Standard Model (SM) signal and one of the most important background
processes to many Higgs-boson production channels.

The cross sections for $W$ and $Z$ boson production with bottom quarks
has been measured at the Tevatron $p \bar p$ collider at Fermilab
($\sqrt{s}=1.96$ TeV) by both the
CDF~\cite{Aaltonen:2008mt,D0Wbb:2008} and
D0~\cite{Abazov:2004jy,Abazov:2004zd} collaborations. These
measurements will continue with increased precision, which will
provide a unique opportunity to test and improve the theoretical
description of heavy-quark jets at hadron colliders by performing a
thorough comparison between the Tevatron experimental data and
existing theoretical predictions. Studying the same cross sections in
the very different kinematic regimes available at the LHC $pp$
collider will then be of great interest and will represent a crucial
test of our understanding of QCD at high-energy colliders.

Moreover, the production of a $W$ and $Z$ boson with one or two
$b$ jets represents a crucial irreducible background for several
Higgs-boson production channels at the LHC.  $Wb\bar{b}$ production is
an irreducible background to $WH$ associated production followed by
the decay $H\rightarrow b\bar{b}$. This is a difficult channel but
theoretically very interesting, since it can play a very important
role in measuring the $b$-quark Yukawa coupling for a light Higgs
boson at the LHC. Analogously, for a light Higgs boson, $Zb\bar{b}$
production is a background to $ZH$ associated production followed by
the decay $H\rightarrow b\bar{b}$. More importantly, for heavier Higgs
bosons, if the $b$ quarks in $Zb\bar{b}$ production decay
leptonically, the $Zb\bar{b}$ production process can be a background
to the inclusive production of a Higgs boson followed by the decay
$H\rightarrow ZZ$ with each $Z$ boson decaying leptonically.  Finally,
$Zb\bar{b}$ production is a background to searches for Higgs bosons
with enhanced $b$-quark Yukawa couplings, produced in $Hb\bar{b}$
associated production followed by the decay $H\rightarrow \mu^+\mu^-$
or $H\rightarrow\tau^+\tau^-$~\cite{Kao:2007qw}.

All Higgs-boson production channels have been calculated including at
least next-to-leading order (NLO) QCD corrections (see, e.~g.,
Ref.~\cite{:2008uu} for a recent review).  The hadronic cross sections
for $gg\rightarrow H$ and associated $WH$ and $ZH$ production are also
known at next-to-next-to-leading order (NNLO) in QCD,
Refs.~\cite{Harlander:2000mg,Harlander:2001is,Harlander:2002wh,
  Anastasiou:2002yz,Anastasiou:2004xq,Ravindran:2003um,
  Catani:2001ic,Catani:2003zt,Bozzi:2005wk,Anastasiou:2008tj,deFlorian:2009hc} and
Refs.~\cite{Han:1991ia,Mrenna:1997wp,Brein:2003wg}, respectively.  The
NLO electroweak corrections to these processes have been calculated as
well (see Ref.~\cite{Actis:2008ug} and references therein
($gg\rightarrow H$) and Ref.~\cite{Ciccolini:2003jy} ($WH,ZH$)).  The
cross section for $Hb\bar{b}$ associated production is known at NLO in
QCD including full $b$-quark mass
effects~\cite{Dawson:2003kb,Dittmaier:2003ej,Dawson:2004sh,Dawson:2004wq}.

The production of a $W$ or a $Z$ boson with up to two jets, one of
which is a $b$ jet, has been calculated including NLO QCD corrections
in the variable-flavor scheme
(VFS)~\cite{Campbell:2003dd,Campbell:2005zv,Campbell:2006cu}, while
the production of a $W$ or $Z$ boson with two $b$ jets has been
derived at NLO in QCD using the fixed-flavor scheme (FFS), first in
the massless $b$-quark
approximation~\cite{Bern:1997sc,Bern:1996ka,Ellis:1998fv,Campbell:2000bg,
  Campbell:2002tg,Campbell:2003hd} and more recently including full
$b$-quark mass
effects~\cite{FebresCordero:2006sj,FebresCordero:2008ci,Cordero:2008ce}.
In the FFS only massless-quark densities are considered in the initial
state, hence the alternative name of 4-flavor-number scheme (4FNS),
while in a VFS an initial-state $b$-quark density is introduced, hence
the alternative name of 5-flavor-number scheme (5FNS). The two schemes
amount to a different ordering of the perturbative series for the
production cross section: in the 4FNS the perturbative series is
ordered strictly by powers of the strong coupling $\alpha_s$, whereas
in the 5FNS the introduction of a $b$-quark parton distribution
function (PDF) allows to resum terms of the form $\alpha_s^n
\ln(m_b^2/M^2)^m$ at all orders in $\alpha_s$ (for fixed order of
logarithms $m$), where $M$ represents the upper integration limit of
the $b$-quark transverse momentum and can be thought to be of the
order of $M_W$ or $M_Z$.  While the two approaches can give very
different results at the leading or lowest order (LO) in QCD, starting
at NLO in QCD they have been shown to be consistent within their
respective theoretical uncertainties for both $H+1\,b$-jet
production~\cite{Campbell:2004pu,Assamagan:2004mu,Kramer:2004ie} (for
a brief review see also Ref.~\cite{Dawson:2004wq}) and single-top
production~\cite{Campbell:2009ss}. Recently, $W+1\,b$-jet production
has been calculated by consistently combining both NLO 4FNS and 5FNS
calculations~\cite{Campbell:2008hh}. Since the relevance of the
logarithms resummed in the VFS approach varies with the kinematic
regime considered, combining the two calculations improves the
accuracy of the theoretical prediction for $W+1\,b$-jet production. A
similar study is currently in progress for $Z+1\,b$-jet
production~\cite{Zb_inprep}. Improving the predictions for
$Z+1\,b$-jet production will be particularly relevant at the LHC,
where this process allows for a direct determination of the $b$-quark
PDF, to be used in the prediction of $H+1\,b$-jet production, a
discovery channel for beyond-the-SM Higgs bosons with enhanced
$b$-quark Yukawa couplings.

In this paper we provide results for $Wb\bar{b}$ and $Zb\bar{b}$
production at the LHC, keeping the $W$ and $Z$ boson on shell and with both $b$ jets tagged in the final state,
i.~e. we focus on the $W/Z+2\,b$-jet case. We include NLO QCD
corrections and full $b$-quark mass effects. The corresponding results
for $Wb\bar{b}$ and $Zb\bar{b}$ production at the Tevatron have been
presented in
Refs.~\cite{FebresCordero:2006sj,FebresCordero:2008ci,Cordero:2008ce}.
The details of the NLO QCD calculations used to compute the NLO cross
sections in this paper can be found in
Refs.~\cite{FebresCordero:2006sj,FebresCordero:2008ci,Cordero:2008ce}.
In this paper we focus on presenting results for the total production
cross sections and kinematic distributions that are of interest to LHC
physics.  Using the MCFM package~\cite{MCFM:2004}, we compare our
results to those obtained by neglecting the $b$-quark mass at NLO in
QCD. A non-zero $b$-quark mass mainly affects phase space regions
where the relevant kinematic observable, such as the transverse
momentum of the bottom quarks or the invariant mass of the $b$-quark
pair, are of the order of $m_b$.  Indeed, this is the reason why
$b$-quark mass effects cannot be neglected in studies of $W/Z+n$-jet
production ($n=1,2$) with at least one $b$ jet, as has been discussed
for the $n=2$ case in Refs.~\cite{Campbell:2005zv,Campbell:2006cu} and
for the $n=1$ case in~\cite{Campbell:2008hh}.

The paper is organized as follows: in Section~\ref{sec:setup} we
briefly describe our choice of input parameters, cuts,
jet-identification algorithm, and observables, while we present
numerical results and a discussion of the $b$-quark mass effects for
$Wb\bar b$ and $Zb\bar b$ production in Section~\ref{sec:wbb} and
Section~\ref{sec:zbb}, respectively.  Section~\ref{sec:conclusions}
contains our conclusions.

\section{General setup}
\label{sec:setup}

The results for both $Wb\bar{b}$ and $Zb\bar{b}$ production presented
in this paper have been obtained for the LHC $pp$ collider running at
a center-of-mass energy of either $\sqrt{s}=10$~TeV or
$\sqrt{s}=14$~TeV.  While we only provide kinematic distributions
obtained with a center-of-mass energy of $\sqrt{s}=14$~TeV, we compare
the total cross sections obtained for center-of-mass energies of
$\sqrt{s}=10$~TeV and $\sqrt{s}=14$~TeV in
Tables~\ref{tab:summary_wbb} and \ref{tab:summary_zbb}. The mass of
the bottom quark is taken to be $m_b=4.62$~GeV. Results in the
massless $b$-quark approximation have been obtained using the MCFM
code (version 5.4) ~\cite{MCFM:2004}. The top-quark mass, entering in
the virtual corrections, is set to $m_t=172.6$~GeV.  In the case of
$Wb\bar{b}$ production we use $M_W=80.44$~GeV, while for $Zb\bar{b}$
production we use $M_Z=91.1876$~GeV and derive $M_W$ from the relation
$M_W=M_Z\cos\theta_W$. In both cases we assume $\sin^2\theta_W=0.223$.
We work in the electroweak $G_\mu$ input scheme and replace the fine
structure constant $\alpha(0)=e^2/(4 \pi)$ by
$\alpha(G_\mu)=\frac{\sqrt{2}}{\pi} G_{\mu} M_W^2 \sin^2\theta_W$ with
the Fermi constant $G_\mu=1.16639 \cdot 10^{-5} \, {\rm GeV}^{-2}$.
The $W$-boson coupling to quarks is proportional to the
Cabibbo-Kobayashi-Maskawa (CKM) matrix elements.  We use non-zero CKM
matrix elements for the first two quark generations,
$V_{ud}=V_{cs}=0.974$ and $V_{us}=V_{cd}=0.227$, while we neglect the
contribution of the third generation, since it is suppressed either by
the initial-state quark PDFs or by the corresponding CKM matrix
elements.
 
The LO results are based on the one-loop evolution of $\alpha_s$ and
the CTEQ6L1 set of PDFs~\cite{Lai:1999wy}, with
$\alpha_s^{LO}(M_Z)=0.130$, while the NLO results use the two-loop
evolution of $\alpha_s$ and the CTEQ6M set of PDFs, with
$\alpha_s^{NLO}(M_Z)=0.118$. In the calculation of the parton
luminosity we assume five light flavors in the initial state, but we
have verified that including the $b$-quark PDF has a negligible effect
($<0.1\%$) on the $W/Z b\bar{b}$ cross section. We implement the $k_T$
jet
algorithm~\cite{Catani:1992zp,Catani:1993hr,Ellis:1993tq,Kilgore:1996sq}
with a pseudo-cone size of $R=0.7$ and we recombine the parton momenta
within a jet using the so called covariant
$E$-scheme~\cite{Catani:1993hr}. We checked that our implementation of
the $k_T$ jet algorithm coincides with the one in MCFM.  We require
all events to have a $b$-jet pair in the final state, with a
transverse momentum larger than either $15$~GeV or $25$~GeV, in order
to study the dependence on the $b$-jet transverse-momentum cut.  We
also require that the pseudorapidities of both $b$ jets satisfy
$|\eta^{b,\bar{b}}|<2.5$. We impose the same $p_T$ and $|\eta|$ cuts
also on the extra jet that may arise due to hard non-collinear real
emission of a parton, i.~e. in the processes $W/Zb\bar{b}+g$ or
$W/Zb\bar{b}+q(\bar{q})$. This hard non-collinear extra parton is
treated either \emph{inclusively} or \emph{exclusively}. In the
\emph{inclusive} case we include both two- and three-jet events, while
in the \emph{exclusive} case we require exactly two jets in the
event. Two-jet events consist of a $b$-jet pair that may also include
a final-state light parton (gluon or quark) due to the applied
recombination procedure. On the other hand, three-jet events consist
of events containing a $b$-jet pair plus an extra light jet.  We
notice that, at NLO in QCD, all jets in three-jet events consist of a
single parton.

For both $Wb\bar{b}$ and $Zb\bar{b}$ production we provide results for
the total cross section ($\sigma$) and the following kinematic
distributions: $d\sigma/dp_T^{b,l}$, $d\sigma/dp_T^{b,sl}$,
$d\sigma/dp_T^{W/Z}$, $d\sigma/d\eta^{b,l}$, $d\sigma/d\eta^{b,sl}$,
$d\sigma/d\eta^{W/Z}$, $d\sigma/dm_{b\bar{b}}$, and
$d\sigma/dR^{b\bar{b}}$, where $p_T^{b,l}$, $p_T^{b,sl}$, and
$p_T^{W/Z}$ are the transverse momenta of the leading $b$ jet
(i.~e. leading in $p_T$), the subleading $b$ jet and of the $W$ or $Z$
boson, $\eta^{b,l}$, $\eta^{b,sl}$, and $\eta^{W/Z}$ are the
corresponding pseudorapidities, $m_{b\bar{b}}$ is the invariant mass
of the $b\bar{b}$ pair, and $R_{b\bar{b}}$ is their relative
separation in the pseudorapidity-azimuthal angle plane,
$R_{b\bar{b}}=\sqrt{(\eta^{b,l}-\eta^{b,sl})^2+(\phi^{b,l}-\phi^{b,sl})^2}$.

\begin{figure}[t]
\begin{center}
\includegraphics*[clip,scale=0.6]{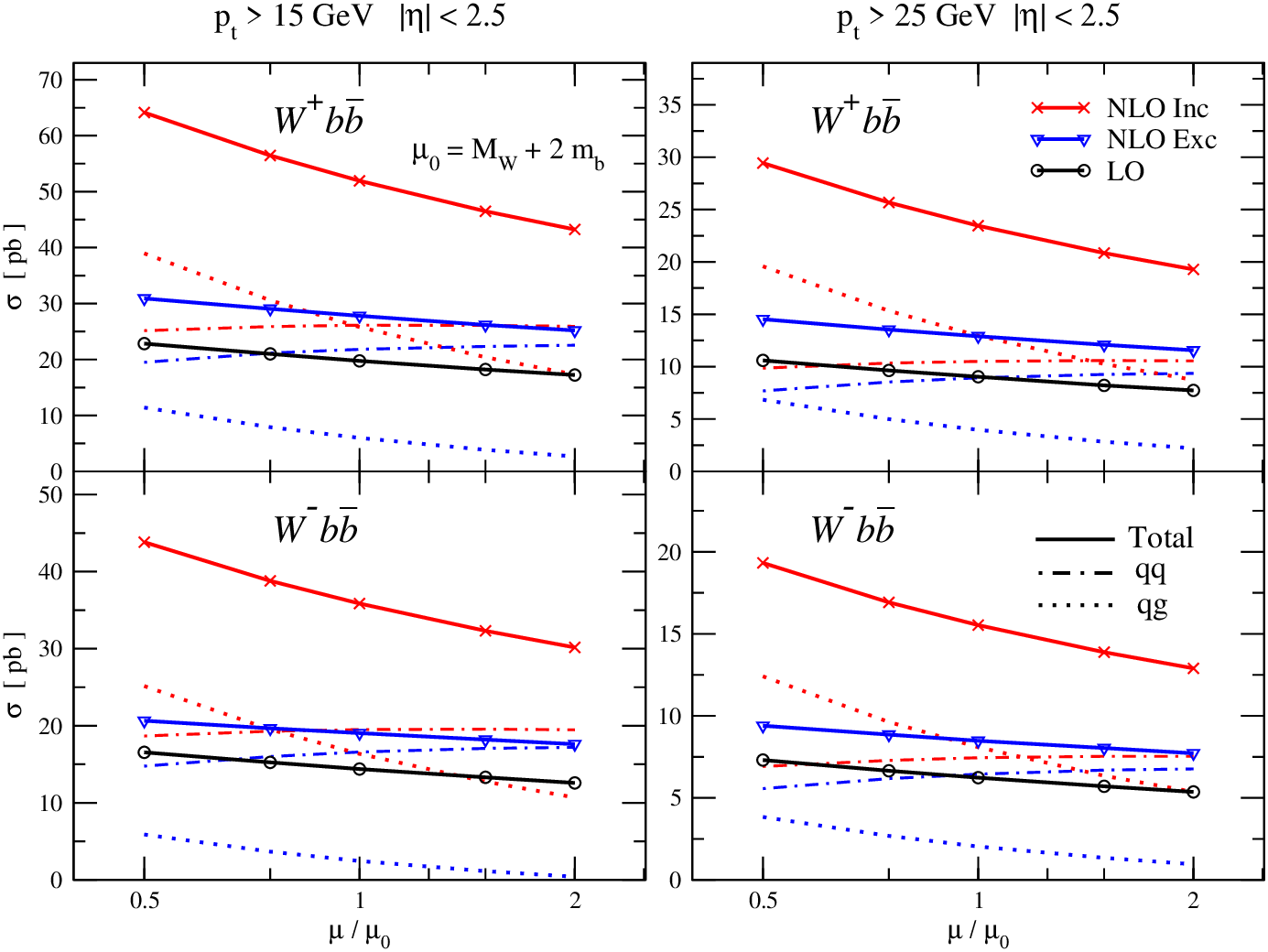} 
\caption[]{Dependence of the LO (black, solid), NLO \emph{exclusive}
  (blue, solid), and NLO \emph{inclusive} (red, solid) total cross
  sections for $W^+b\bar{b}$ and $W^-b\bar{b}$ production on the
  renormalization/factorization scales, including full $b$-quark mass
  effects, when $\mu=\mu_r=\mu_f$ is varied between $\mu_0/2$ and
  $2\mu_0$ (with $\mu_0=M_W+2m_b$). We also show the individual
  channels, $q\bar{q}^\prime$ (dash-dotted) and $qg+\bar{q}g$
  (dotted), for the \emph{inclusive} (red) and \emph{exclusive}
  (blue) cases.}
\label{fig:Wbb_mudep}
\end{center}
\end{figure}

\begin{table}[htb]
\begin{center}
  \caption{LO, NLO \emph{inclusive}, and NLO \emph{exclusive} cross
    sections for $W^+b\bar{b}$ and $W^-b\bar{b}$ production at both
    $\sqrt{s}=10$~TeV and $\sqrt{s}=14$~TeV, for two different values
    of the $b$-jet transverse-momentum selection cut, and for a zero
    (obtained with MCFM) and non-zero $b$-quark mass. The central
    values correspond to $\mu_r=\mu_f=\mu_0=M_W+2m_b$, while the upper
    and lower bounds represent the maximal upper and lower variation
    obtained when varying $\mu_r=\mu_f$ between $\mu_0/2$ and
    $2\mu_0$.}
\label{tab:summary_wbb}
\begin{tabular}{|c|c|c|c|c|c|c|c|c|} \hline\hline
 & \multicolumn{4}{|c|}{$W^+b\bar{b}$} &
\multicolumn{4}{|c|}{$W^-b\bar{b}$}\\
& \multicolumn{2}{|c|}{$p_T^b>15$~GeV} & \multicolumn{2}{|c|}{$p_T^b>25$~GeV} & \multicolumn{2}{|c|}{$p_T^b>15$~GeV} &\multicolumn{2}{|c|}{$p_T^b>25$~GeV} \\ 
\hline
& $m_b\neq 0$ & $m_b=0$ & $m_b\neq 0$ & $m_b=0$ & $m_b\neq 0$ & $m_b=0$ & $m_b\neq 0$ & $m_b=0$ \\ 
\hline
&\multicolumn{8}{|c|}{$\sqrt{s}=10$~TeV} \\
\hline
$\sigma_{LO}$ (pb) & $14.4^{+2.6}_{-2.1}$ & $15.5^{+2.7}_{-2.2}$ & $6.49^{+1.3}_{-1.0}$ & $6.67^{+1.3}_{-1.1}$ & $9.77^{+1.7}_{-1.4}$ & $10.5^{+1.8}_{-1.5}$ & $4.15^{+0.83}_{-0.66}$ & 
$4.27^{+0.83}_{-0.68}$ \\
$\sigma_{NLO,inc}$ (pb) & $33.6^{+7.8}_{-5.6}$ & $36.4^{+8.1}_{-6.2}$ & $14.6^{+3.6}_{-2.5}$ & $15.1^{+3.6}_{-2.7}$ & $22.1^{+4.9}_{-3.5}$ & $24.0^{+5.2}_{-3.9}$ & $9.16^{+2.2}_{-1.5}$ & 
$9.49^{+2.1}_{-1.7}$ \\
$\sigma_{NLO,exc}$ (pb) & $18.6^{+1.7}_{-1.6}$ & $20.3^{+1.7}_{-1.8}$ & $8.37^{+0.84}_{-0.77}$ & $8.67^{+0.85}_{-0.87}$ & $12.3^{+0.86}_{-0.90}$ & $13.4^{+0.9}_{-1.2}$ & $5.29^{+0.48}_{-0.45}$ & 
$5.50^{+0.43}_{-0.51}$ \\
\hline
&\multicolumn{8}{|c|}{$\sqrt{s}=14$~TeV} \\
\hline
$\sigma_{LO}$ (pb) & $19.8^{+3.1}_{-2.5}$ & $21.3^{+3.2}_{-2.7} $ & $9.02^{+1.6}_{-1.3}$ & $9.26^{+1.6}_{-1.3}$ & $14.4^{+2.1}_{-1.8}$ & $15.5^{+2.2}_{-2.0}$ & $6.24^{+1.1}_{-0.87}$ & $6.42^{+1.1}_{-0.91}$ \\
$\sigma_{NLO,inc}$ (pb) & $51.9^{+12}_{-8.7}$ & $56.3^{+13}_{-9.6}$ & $23.4^{+6.0}_{-4.2}$ & 
$24.3^{+5.9}_{-4.5}$ & $35.8^{+7.9}_{-5.7}$ & $39.0^{+8.5}_{-6.4}$ & $15.5^{+3.8}_{-2.6}$ & 
$16.1^{+3.7}_{-2.8}$ \\
$\sigma_{NLO,exc}$ (pb) & $27.8^{+3.1}_{-2.5}$ & $30.4^{+3.5}_{-2.8}$ & $12.9^{+1.6}_{-1.3}$ & 
$13.4^{+1.6}_{-1.5}$ & $19.0^{+1.6}_{-1.4}$ & $20.8^{+1.7}_{-1.5}$ & $8.49^{+0.90}_{-0.77}$ & 
$8.83^{+0.93}_{-0.81}$ \\
\hline
\end{tabular}
\end{center}
\end{table}

\begin{figure}[htb]
\begin{center}
\begin{tabular}{l}
\includegraphics*[clip,scale=0.5]{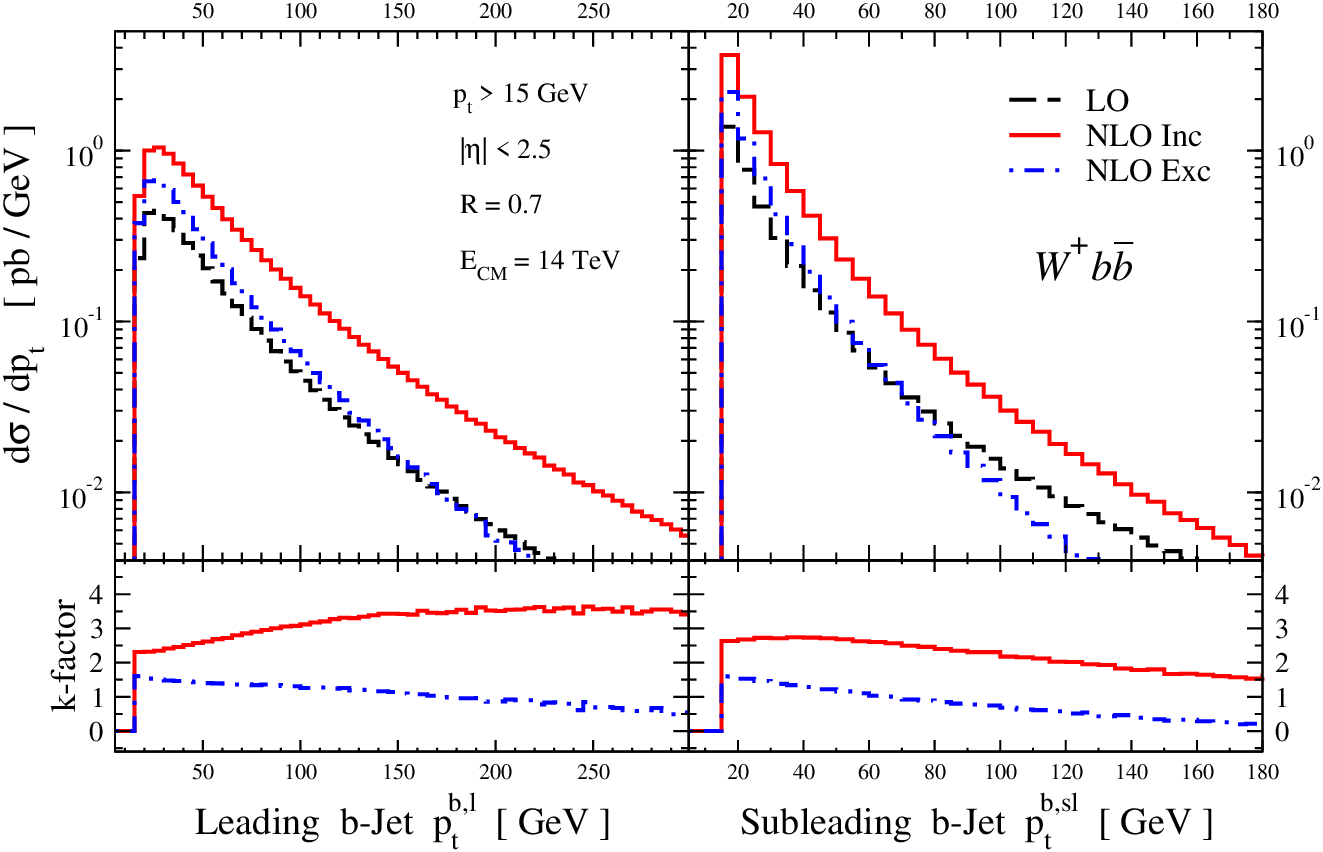} \\
\includegraphics*[clip,scale=0.5]{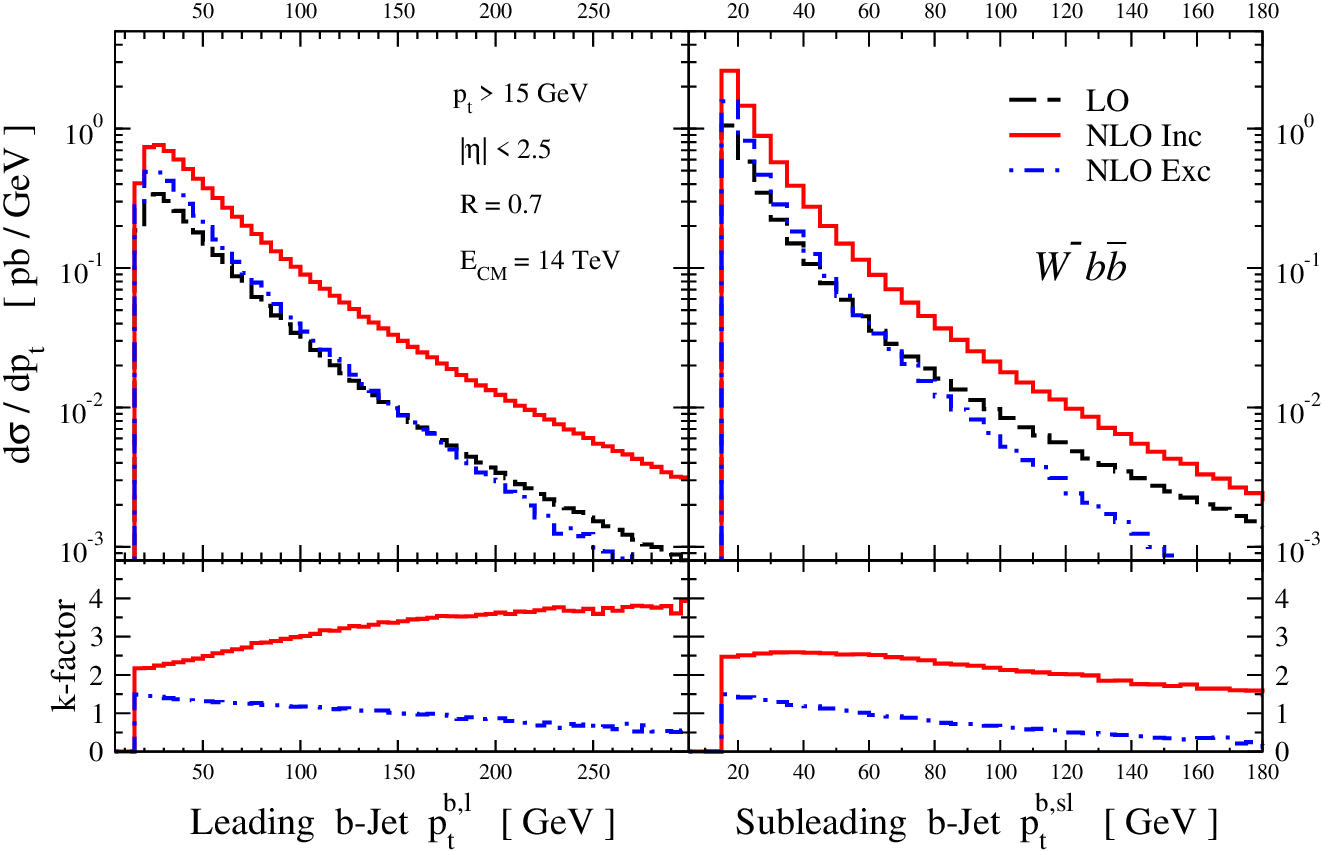} 
\end{tabular}
\caption[]{LO (black, dashed), NLO \emph{inclusive} (red, solid) and
  NLO \emph{exclusive} (blue, dot-dashed) transverse momentum
  distributions for the $b$ jet with the leading (left hand side) and
  subleading (right hand side) transverse momentum in $W^+b\bar{b}$
  (upper plots) and $W^-b\bar{b}$ (lower plots) production. The lower
  window shows a bin-by-bin K factor, for the \emph{inclusive} (red,
  solid) and \emph{exclusive} (blue, dot-dashed) cases.}
\label{fig:Wbb_pt_bjets_pt15}
\end{center}
\end{figure}
\begin{figure}[htb]
\begin{center}
\begin{tabular}{ll}
\includegraphics*[clip,scale=0.46]{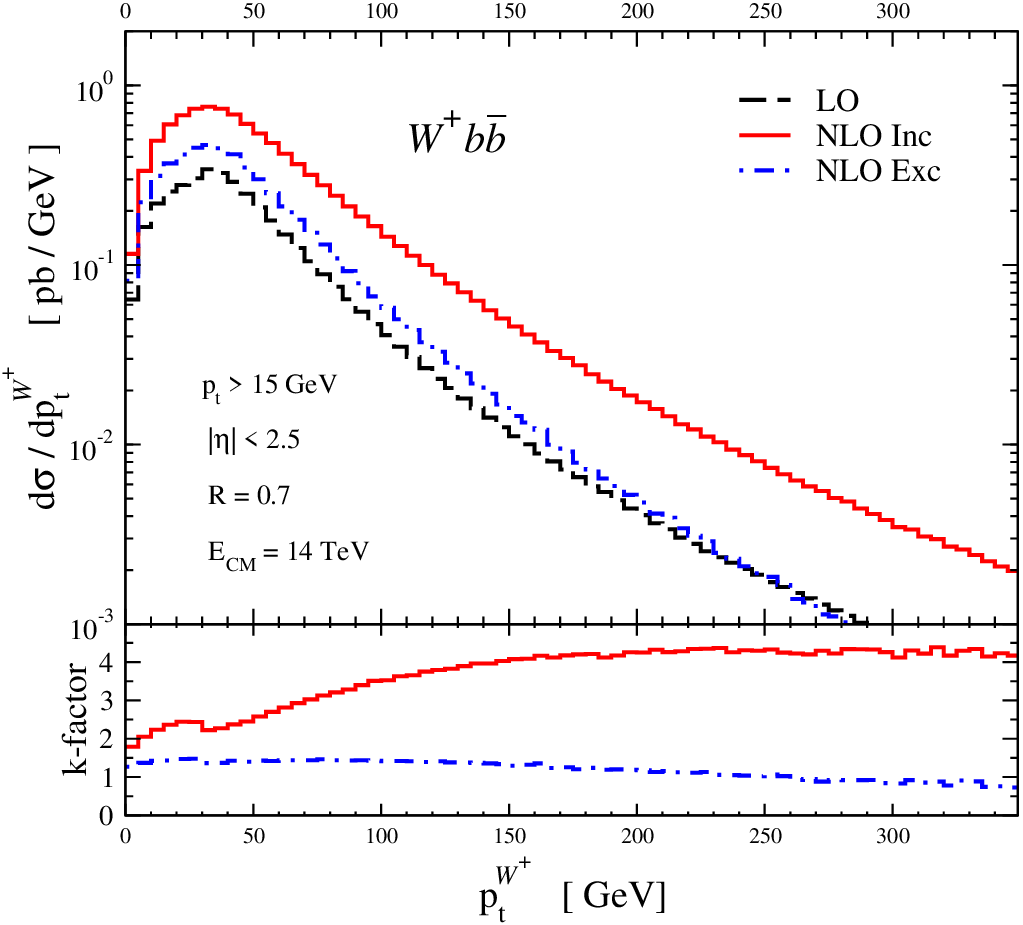} &
\includegraphics*[clip,scale=0.46]{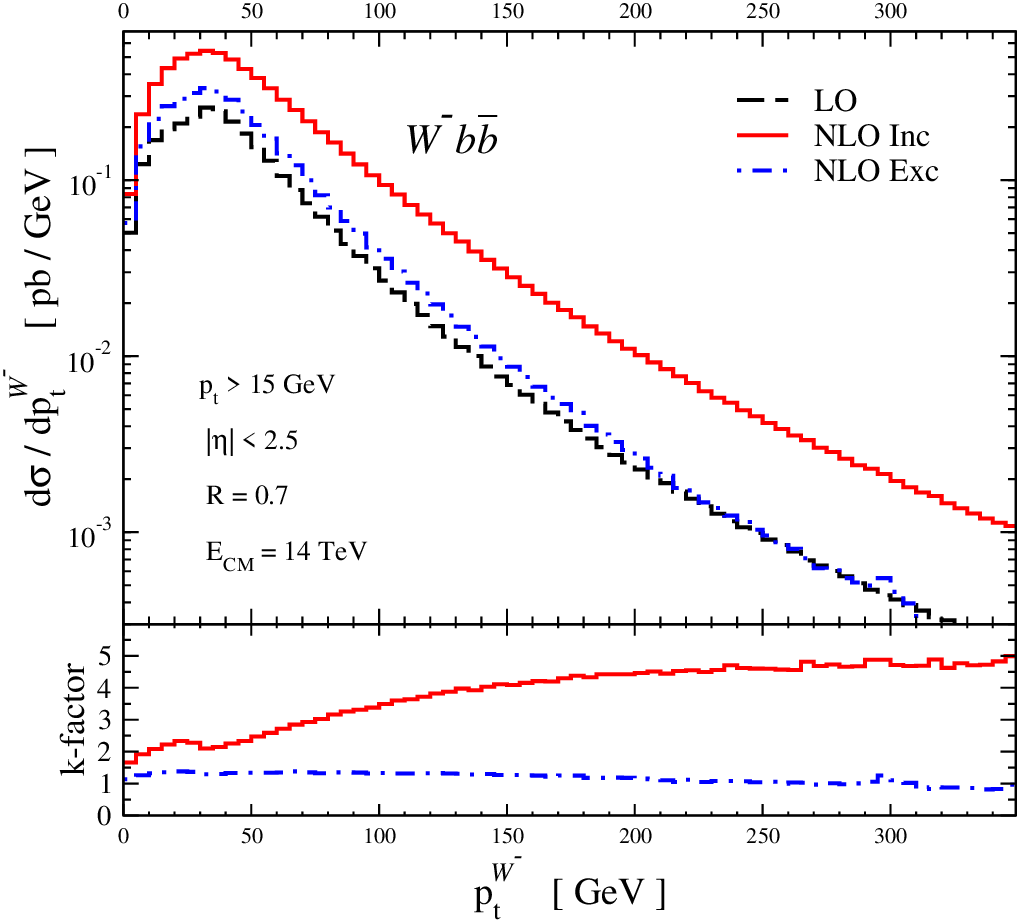} 
\end{tabular}
\caption[]{LO (black, dashed), NLO \emph{inclusive} (red, solid) and
  NLO \emph{exclusive} (blue, dot-dashed) transverse momentum
  distributions for the $W^+$ (left hand side) and $W^-$ (right hand
  side) bosons in $W^+b\bar{b}$ and $W^-b\bar{b}$ production
  respectively. The lower window shows a bin-by-bin K factor, for the
  \emph{inclusive} (red, solid) and \emph{exclusive} (blue,
  dot-dashed) cases.}
\label{fig:Wbb_pt_W_pt15}
\end{center}
\end{figure}
\begin{figure}[htb]
\begin{center}
\begin{tabular}{l}
\includegraphics*[clip,scale=0.5]{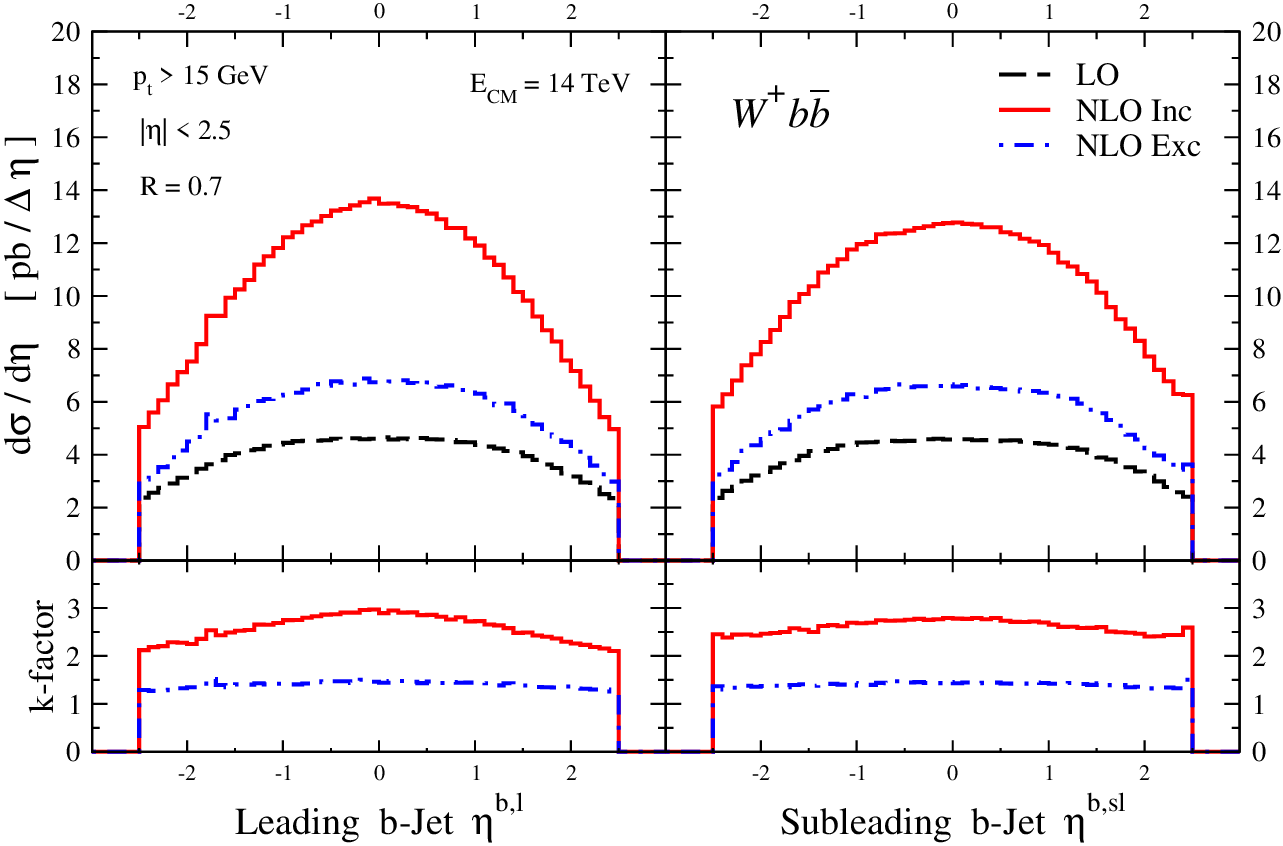}\\
\includegraphics*[clip,scale=0.5]{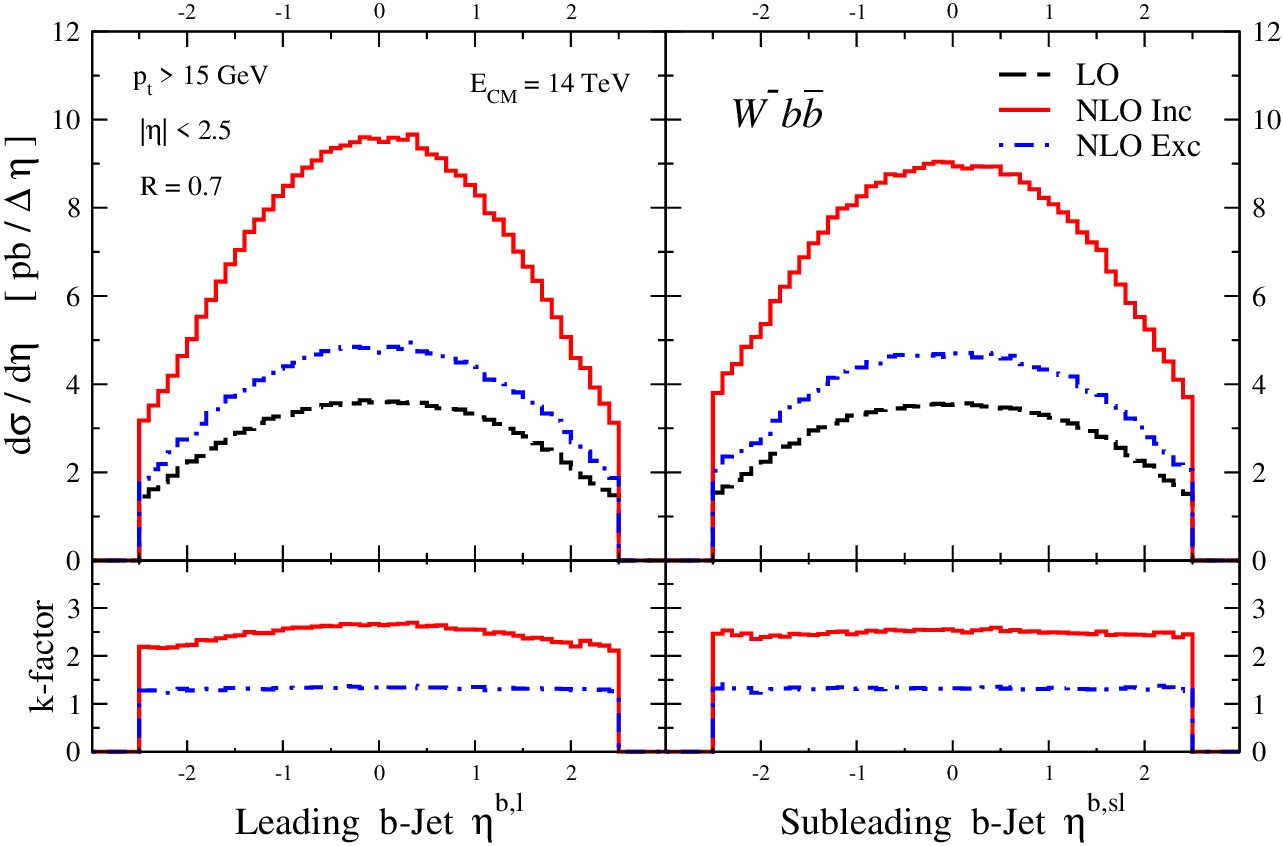} 
\end{tabular}
\caption[]{LO (black, dashed), NLO \emph{inclusive} (red, solid) and
  NLO \emph{exclusive} (blue, dot-dashed) pseudorapidity distributions
  for the $b$ jet with the leading (left hand side) and subleading
  (right hand side) transverse momentum in $W^+b\bar{b}$ (upper plots)
  and $W^-b\bar{b}$ (lower plots) production. The lower window shows a
  bin-by-bin K factor, for the \emph{inclusive} (red, solid) and
  \emph{exclusive} (blue, dot-dashed) cases.}
\label{fig:Wbb_eta_bjets_pt15}
\end{center}
\end{figure}
\begin{figure}[htb]
\begin{center}
\begin{tabular}{ll}
\includegraphics*[clip,scale=0.46]{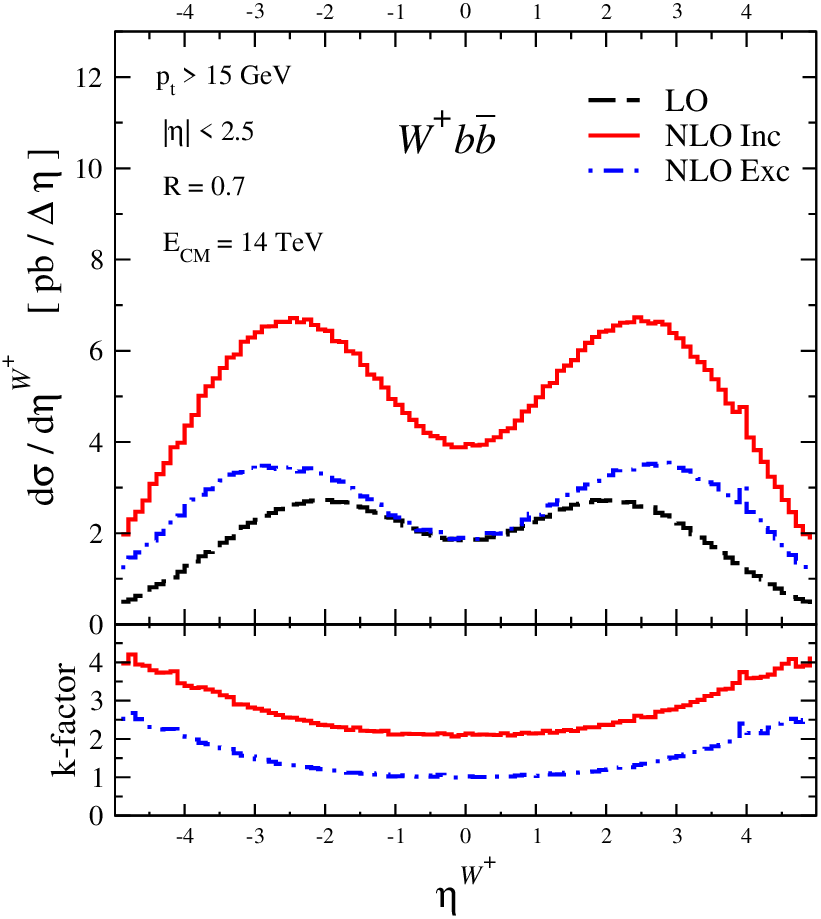} &
\includegraphics*[clip,scale=0.46]{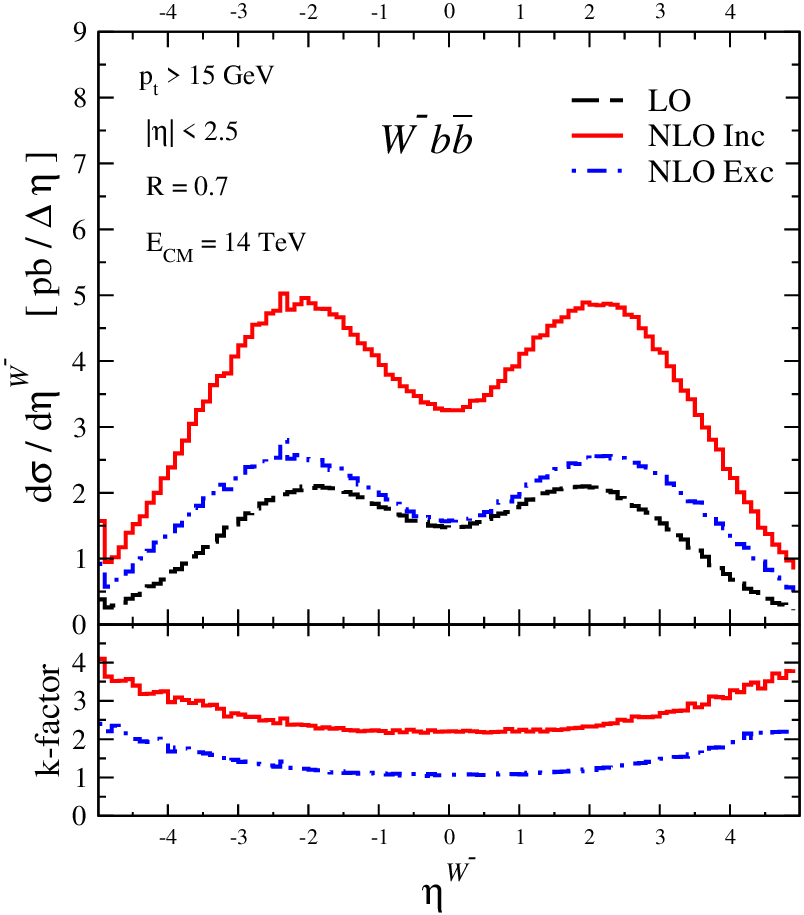} 
\end{tabular}
\caption[]{LO (black, dashed), NLO \emph{inclusive} (red, solid) and
  NLO \emph{exclusive} (blue, dot-dashed) pseudorapidity distributions
  for the $W^+$ (left hand side) and $W^-$ (right hand side) bosons in
  $W^+b\bar{b}$ and $W^-b\bar{b}$ production respectively. The lower
  window shows a bin-by-bin K factor, for the \emph{inclusive} (red,
  solid) and \emph{exclusive} (blue, dot-dashed) cases.}
\label{fig:Wbb_eta_W_pt15}
\end{center}
\end{figure}
\begin{figure}[htb]
\begin{center}
\begin{tabular}{ll}
\includegraphics*[clip,scale=0.46]{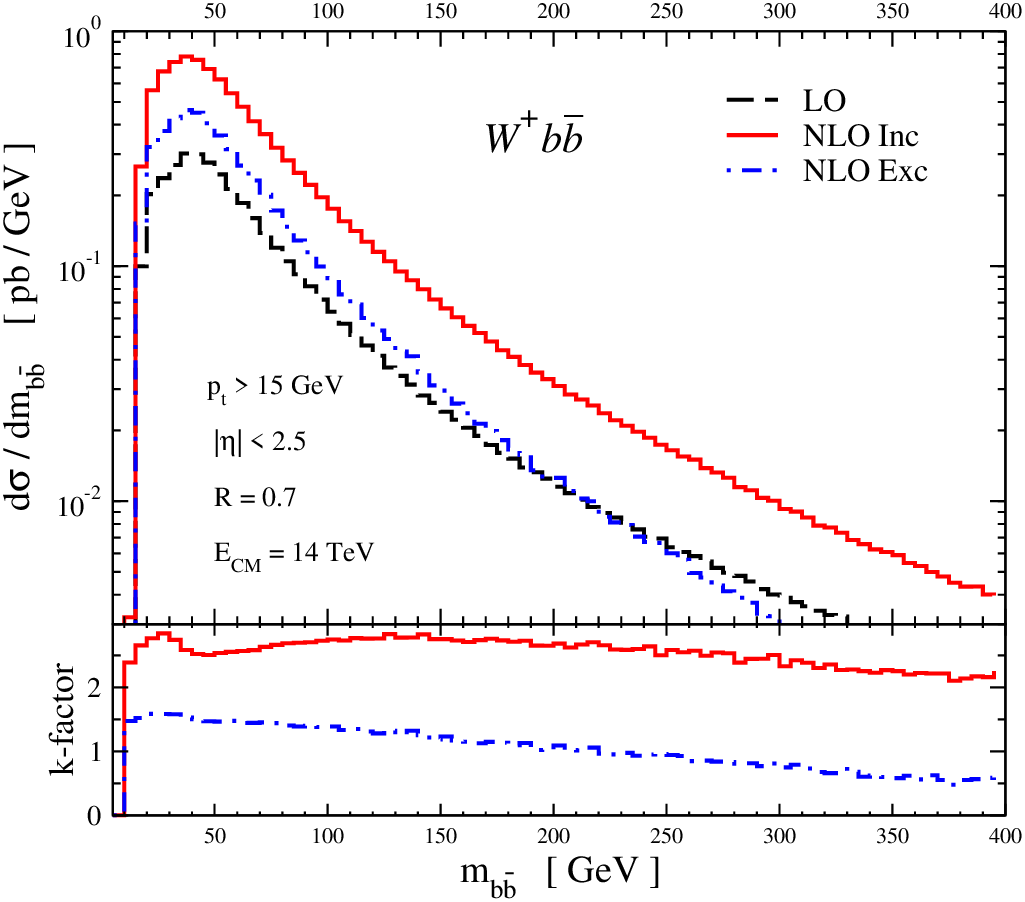} &
\includegraphics*[clip,scale=0.46]{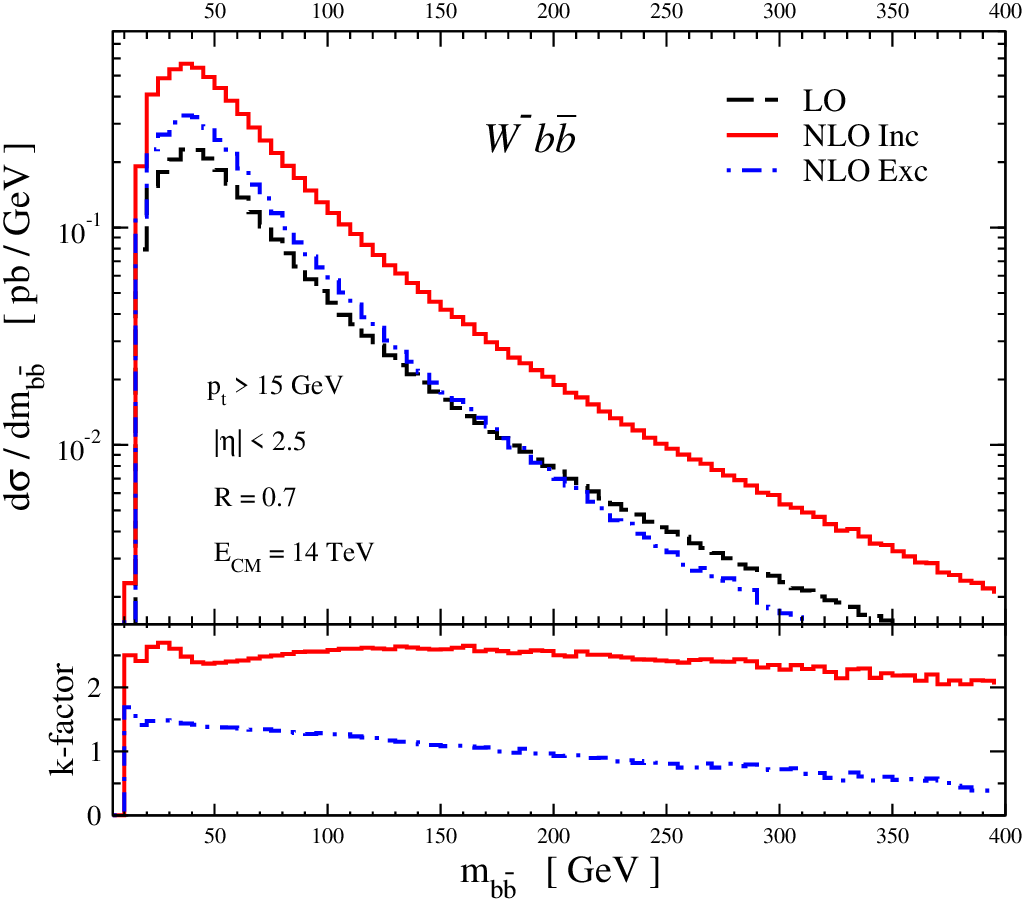} 
\end{tabular}
\caption[]{LO (black, dashed), NLO \emph{inclusive} (red, solid) and
  NLO \emph{exclusive} (blue, dot-dashed) $b\bar{b}$-pair invariant
  mass distributions in $W^+b\bar{b}$ (left hand side) and
  $W^-b\bar{b}$ (right hand side) production. The lower window shows a
  bin-by-bin K factor, for the \emph{inclusive} (red, solid) and
  \emph{exclusive} (blue, dot-dashed) cases.}
\label{fig:Wbb_mbb_pt15}
\end{center}
\end{figure}
\begin{figure}[htb]
\begin{center}
\begin{tabular}{ll}
\includegraphics*[clip,scale=0.46]{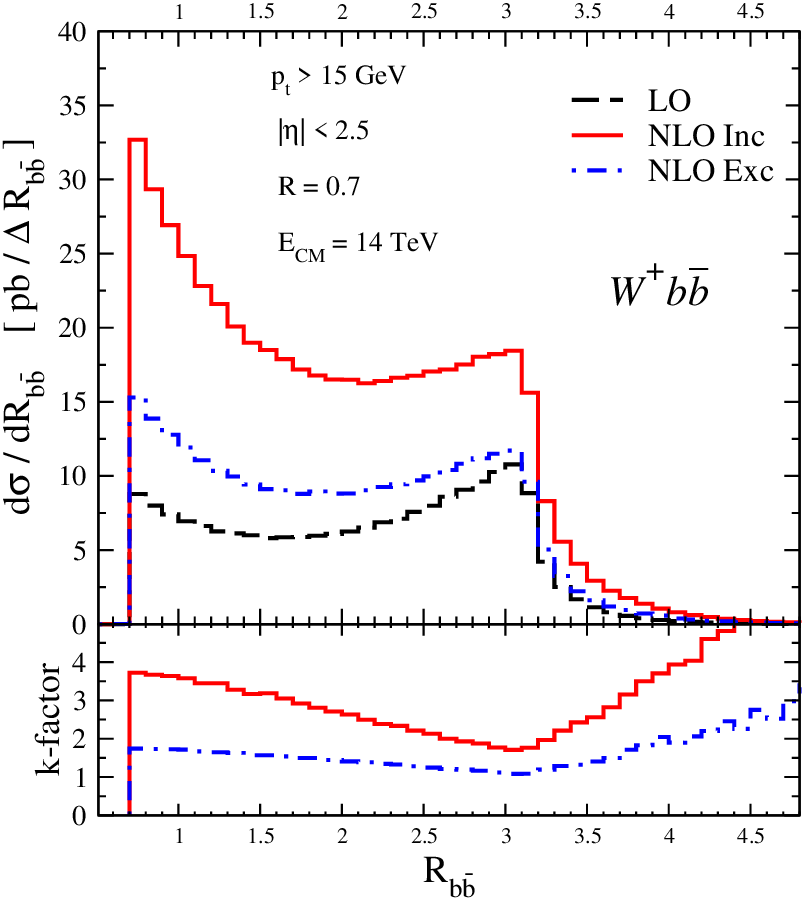} &
\includegraphics*[clip,scale=0.46]{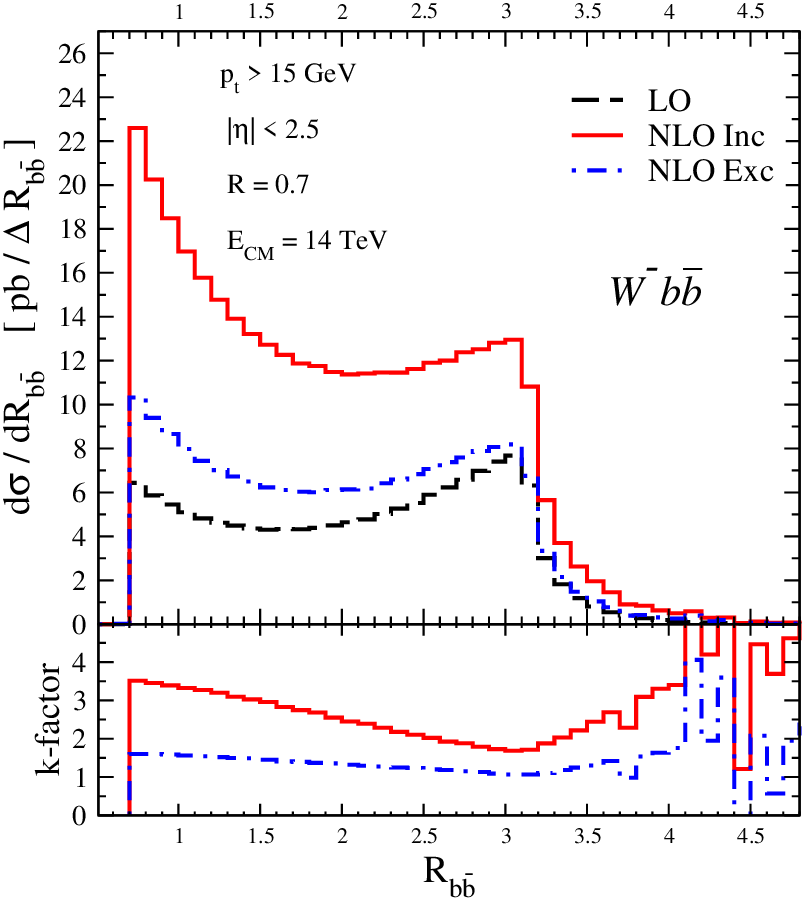} 
\end{tabular} 
\caption[]{LO (black, dashed), NLO \emph{inclusive} (red, solid) and
  NLO \emph{exclusive} (blue, dot-dashed) $R_{b\bar b}$ distributions
  in $W^+b\bar{b}$ (left hand side) and $W^-b\bar{b}$ (right hand
  side) production. The lower window shows a bin-by-bin K factor, for
  the \emph{inclusive} (red, solid) and \emph{exclusive} (blue,
  dot-dashed) cases.}
\label{fig:Wbb_Rbb_pt15}
\end{center}
\end{figure}
\begin{figure}[htb]
\begin{center}
\begin{tabular}{ll}
\includegraphics*[clip,scale=0.52]{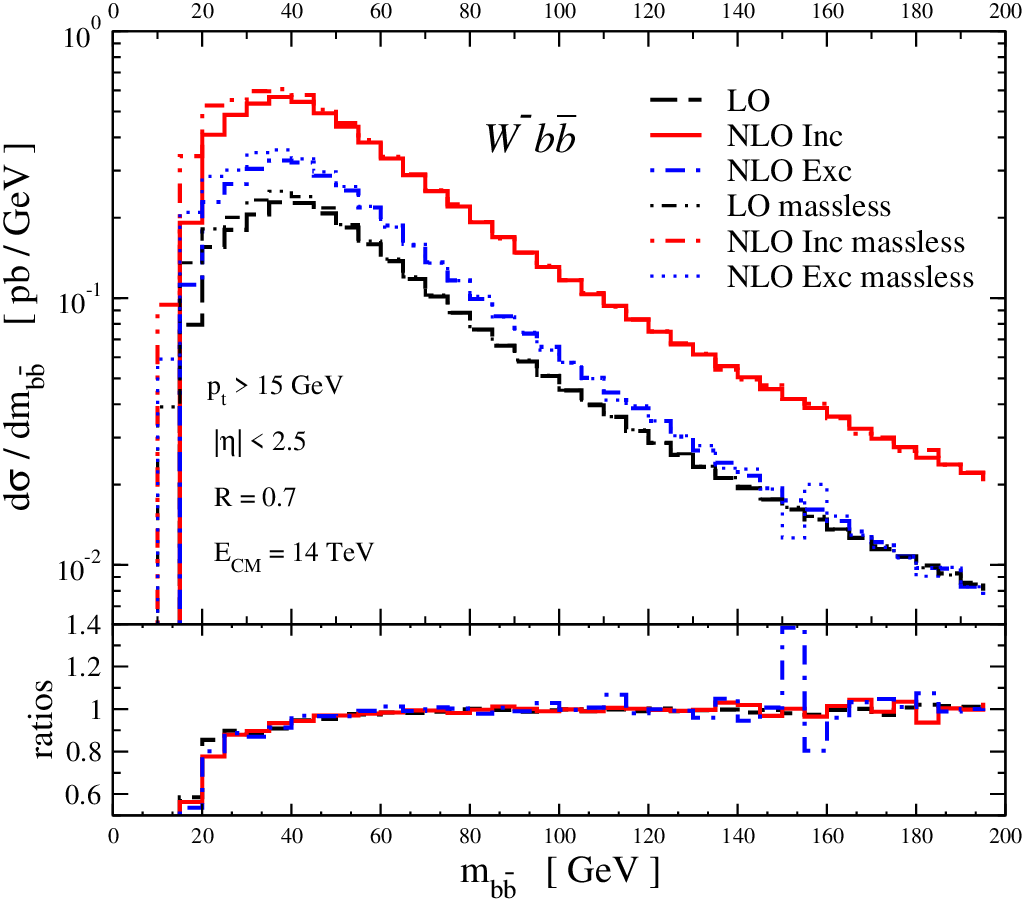} &
\includegraphics*[clip,scale=0.52]{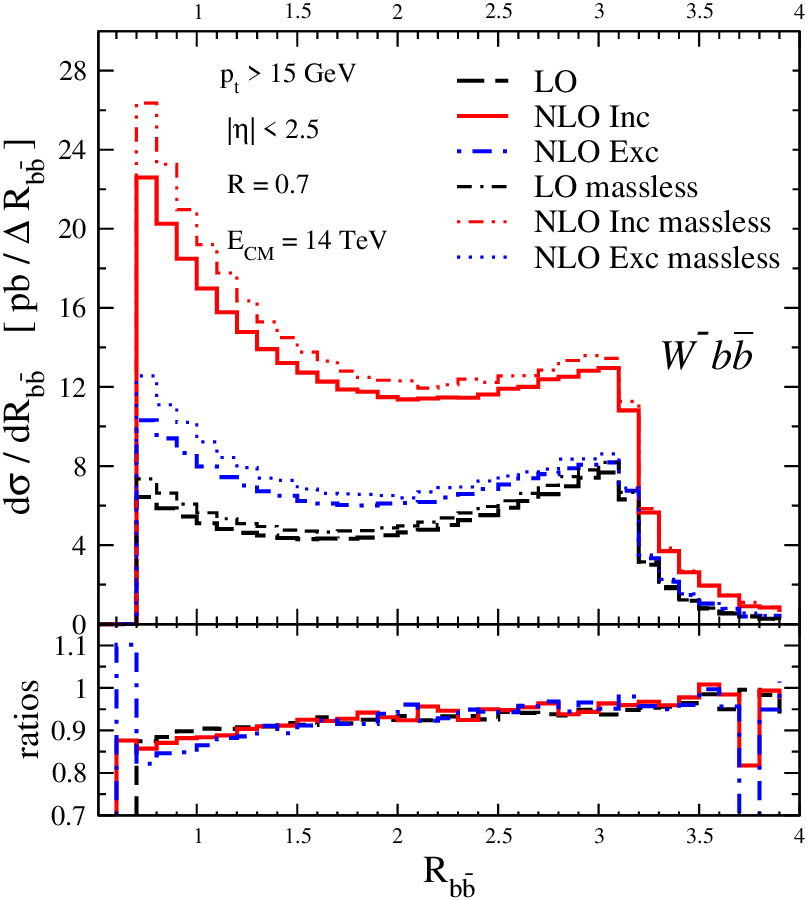} 
\end{tabular} 
\caption[]{LO (black), NLO \emph{inclusive} (red) and NLO
  \emph{exclusive} (blue) $m_{b\bar{b}}$ (left hand side) and
  $R_{b\bar{b}}$ (right hand side) distributions for $W^-b\bar{b}$
  production derived from our calculation with $m_b\ne 0$ (LO: dashed,
  NLO \emph{inclusive}: solid, NLO \emph{exclusive}: dash-dotted) and
  from MCFM with $m_b=0$ (LO: double-dashed/dotted, NLO
  \emph{inclusive}: dashed/double-dotted, NLO \emph{exclusive}:
  dotted).  The lower window shows the ratio of the distributions for
  massive and massless $b$ quarks, $d\sigma(m_b\neq 0)/d\sigma(m_b=0)$
  (LO: dashed, NLO \emph{inclusive}: solid, NLO \emph{exclusive}:
  dash-dotted).}
\label{fig:Wbb_comp_pt15}
\end{center}
\end{figure}

\boldmath
\section{$Wb\bar{b}$ production}
\label{sec:wbb}
\unboldmath 

At tree level, the production of a $W$ boson with a pair of bottom
quarks consists of just one process, $q\bar{q}^\prime\rightarrow
Wb\bar{b}$.  In order to compute this process at NLO in QCD one needs
to include one-loop virtual corrections to $q\bar{q}^\prime\rightarrow
Wb\bar{b}$ as well as all real radiation corrections with up to one
extra parton in the final state, i.~e.  $q\bar{q}^\prime\rightarrow
Wb\bar{b}+g$ and $qg(\bar{q}g)\rightarrow
Wb\bar{b}+q^\prime(\bar{q}^\prime)$.  Details of the calculation have
been given in Refs.~\cite{FebresCordero:2006sj,Cordero:2008ce} and
will not be repeated here.

We note that, contrary to a $p\bar{p}$ collider like the Tevatron, at
a $pp$ collider like the LHC the cross sections for $W^+b\bar{b}$ and
$W^-b\bar{b}$ are different, because the two processes depend on
different quark/antiquark PDFs that are not symmetrically distributed
between the two incoming nucleons.  In the following we will provide
results separately for both production processes.

In Fig.~\ref{fig:Wbb_mudep} we illustrate the renormalization- and
factorization-scale dependence of the LO and NLO total cross sections
obtained for a massive $b$ quark, when $\mu=\mu_r=\mu_f$ is varied
between $\mu_0/2$ and $2\mu_0$, with $\mu_0=M_W+2m_b$. We immediately
notice that the impact of NLO QCD corrections is very large, in
particular in the \emph{inclusive} case, where they increase the LO
cross section by a factor between two and three depending on the
scale. We also notice that the scale dependence of the NLO cross
section is worse than (\emph{inclusive} case) or comparable to
(\emph{exclusive} case) the scale dependence of the LO cross
section. This is different from what has been observed for the
Tevatron~\cite{FebresCordero:2006sj,Cordero:2008ce}, and was first
pointed out in a calculation with massless bottom
quarks~\cite{Campbell:2003hd}.  It is just a reminder of the fact
that, at a given perturbative order, the uncertainty due to the
residual renormalization- and factorization-scale dependence may
underestimate the theoretical uncertainty due to missing higher-order
corrections. A realistic determination of this uncertainty is usually
much more complex and requires a thorough understanding of the
perturbative structure of the cross section, in particular at the
lowest orders of the perturbative expansion.  In $Wb\bar b$ production
the NLO QCD corrections introduce a new, numerically important
production channel not present at LO, as will be discussed in more
detail below.  Therefore, only at NLO the scale dependence of the
cross sections starts to be a meaningful measure of the behavior of
the perturbative expansion.

In order to understand better the behavior of the $W^+b\bar{b}$ and
$W^-b\bar{b}$ cross section, we also show in Fig.~\ref{fig:Wbb_mudep}
the scale dependence of the individual parton-level channels. We
notice that although the $qg(\bar{q}g)\rightarrow
W^{\pm}b\bar{b}+q^\prime(\bar{q}^\prime)$ channel appears for the
first time at NLO in the perturbative expansion of the $Wb\bar{b}$
cross section, it is actually a tree level contribution and, as such,
introduces a large scale dependence in the calculation that will be
moderated only by adding (still unknown) NNLO corrections.  The reason
why this becomes so evident at the LHC, while it is not at the
Tevatron, is because the $qg(\bar{q}g)\rightarrow
W^{\pm}b\bar{b}+q^\prime(\bar{q}^\prime)$ channel is enhanced by the
correspondingly large initial-state gluon PDF. The NLO $Wb\bar{b}$
total cross section is particularly affected by this process because
there is no gluon-initiated process at LO.  Finally, the impact of the
$qg(\bar{q}g)\rightarrow W^{\pm}b\bar{b}+q^\prime(\bar{q}^\prime)$
channel on the scale dependence of the total cross section is larger
in the \emph{inclusive} than in the \emph{exclusive} case because the
\emph{exclusive} cross section by definition discriminates against
processes with more than two jets in the final state.
Figure~\ref{fig:Wbb_mudep} also shows the effect of lowering the cut on
the transverse momentum of the $b$ jets. Lowering the cut from
$p_T^b>25$~GeV to $p_T^b>15$~GeV almost doubles the cross section and
can therefore be a crucial factor in deciding which selection cuts to
use for $b$ jets.

In Table~\ref{tab:summary_wbb} we present both LO and NLO total cross
sections separately for $W^+b\bar{b}$ and $W^-b\bar{b}$ production,
including our estimate of the scale uncertainty due to the residual
renormalization- and factorization-scale dependence. We provide
results for both $p_T^b>15$~GeV and $p_T^b>25$~GeV, and for both
center-of-mass energies, $\sqrt{s}=14$~TeV and $\sqrt{s}=10$~TeV.  We
also include the corresponding set of results obtained with MCFM for
$m_b=0$.  Comparing the results from the massless approximation and
our results with full $b$-quark mass dependence, one observes that for
$p_T^b>15$~GeV the massless approximation overestimates the total
cross section by about $10\%$, while for $p_T^b>25$~GeV the difference
is a milder $3\%$ (both for $\sqrt{s}=10$~TeV and
$\sqrt{s}=14$~TeV). As expected, the more inclusive the treatment of
the $b$ jets the more important the $b$-quark mass effects
become. This, for example, explains why for a complete NLO treatment
of $W+1\, b$-jet production, the contributions from the $Wb\bar b$
production process must be calculated using the full $b$-quark mass
dependence, as discussed in~\cite{Campbell:2008hh}.

In Figs.~(\ref{fig:Wbb_pt_bjets_pt15})-(\ref{fig:Wbb_eta_W_pt15}) we
show the transverse momentum ($p_T$) and pseudorapidity ($\eta$)
distributions for the $b$ jet with the leading and subleading
transverse momentum and for the $W$ boson for both $W^+b\bar b$ and $W^-b \bar b$
production. The results for these distributions, as well as for all
other distributions presented in this paper, have been obtained with
$\sqrt{s}=14$~TeV and by assuming $p_T^b> 15$~GeV and
$|\eta^b|<2.5$. The upper panels of each figure show the LO, NLO
\emph{inclusive} and NLO \emph{exclusive} distributions, while the
lower panels show the ratios $d\sigma_{NLO}^{inc}/d\sigma_{LO}$ and
$d\sigma_{NLO}^{exc}/d\sigma_{LO}$, thereby providing a bin-by-bin
K factor. In each figure we show the results obtained with the central
scale choice $\mu_r=\mu_f=\mu_0=M_W+2m_b$.  The distributions for the
invariant mass of the two $b$ jets ($m_{b\bar{b}}$) and for their
relative separation in the pseudorapidity-azimuthal angle plane
($R_{b\bar{b}}$) are shown in Fig.~(\ref{fig:Wbb_mbb_pt15}) and
(\ref{fig:Wbb_Rbb_pt15}), respectively.

Clearly, the NLO QCD corrections largely affect the kinematic
distributions, resulting in considerable changes in their shapes, both
in the \emph{inclusive} and \emph{exclusive} case.
Figure~(\ref{fig:Wbb_pt_bjets_pt15}) shows that at NLO in QCD the
production of $b$ jets at large $p_T^b$ is consistently suppressed in
the \emph{exclusive} case, while in the \emph{inclusive} case the
cross section for the production of a leading $b$ jet is enhanced,
yielding K factors of about $2$ in the low $p_T^b$ region and of about
$3.5$ at large $p_T^b$. Similar features can be observed in the
$p_T^W$ distributions of Fig.~(\ref{fig:Wbb_pt_W_pt15}). We note
that this may have an impact especially on assessing the effects of
experimental triggers on the lepton coming from the $W$-boson
decay. The pseudorapidity distributions of the leading and sub-leading
$b$ jet of
Figs.~(\ref{fig:Wbb_eta_bjets_pt15}) and (\ref{fig:Wbb_eta_W_pt15}) in the
\emph{exclusive} case are enhanced by the NLO QCD corrections but
their shape is barely affected, while in the \emph{inclusive} case the
increase of the $\eta^b$ distribution of the leading $b$ jet is more
pronounced in the central region. The increase of the pseudorapidity
distributions of the $W$ bosons at NLO QCD, on the other hand, is more
pronounced in the forward regions in both the \emph{inclusive} and
\emph{exclusive} cases. Finally, we point out that the large positive
corrections to the $R_{b \bar b}$ distributions of
Fig.~(\ref{fig:Wbb_Rbb_pt15}) in regions with low and large values
of $R_{b\bar b}$ are especially pronounced in the \emph{inclusive}
case. Although these changes in shape are commonly seen in NLO QCD
computations involving two or more jets, the large effects observed
here in the \emph{inclusive} case are a specific feature of $Wb\bar b$
production, originating from the tree-like $qg$-initiated contribution
to the NLO QCD corrections, as one can easily deduce from comparing
with the \emph{exclusive} K factors.

Finally, we compared the distributions presented here to the ones
produced by MCFM in the massless $b$-quark approximation. We agree
very well, with only small (of the order of the change in the total
cross sections shown in Table~\ref{tab:summary_wbb}) but noticeable
deviations in regions where relevant kinematic observables become
small, i.~e. comparable to $m_b$.  As an example, we show in
Fig.~\ref{fig:Wbb_comp_pt15} the comparison of the LO and NLO
$m_{b\bar{b}}$ and $R_{b\bar{b}}$ distributions for $W^- b\bar b$
production obtained from the massive and massless $b$-quark
calculations.  Most of the difference between the massless and massive
$b$-quark cross sections comes from the region of low invariant mass
$m_{b\bar{b}}$, and are more pronounced for small values of $R_{b\bar
  b}$, both at LO and at NLO, where the cross sections for $m_b\ne 0$
are consistently smaller than the ones with $m_b=0$. This may indicate
that a resummation of large logarithmic corrections may be in order
when the two $b$ jets become collinear. As
can be seen by comparing the ratios of the LO and NLO cross sections
in Fig.~\ref{fig:Wbb_comp_pt15}, the impact of a non-zero $b$-quark
mass is almost not affected by including NLO QCD corrections and can
be taken into account by rescaling the NLO result for massless bottom
quarks with the ratio of the LO cross sections as discussed
in~\cite{FebresCordero:2006sj}.

\begin{figure}[htb]
\begin{center}
\includegraphics*[clip,scale=0.6]{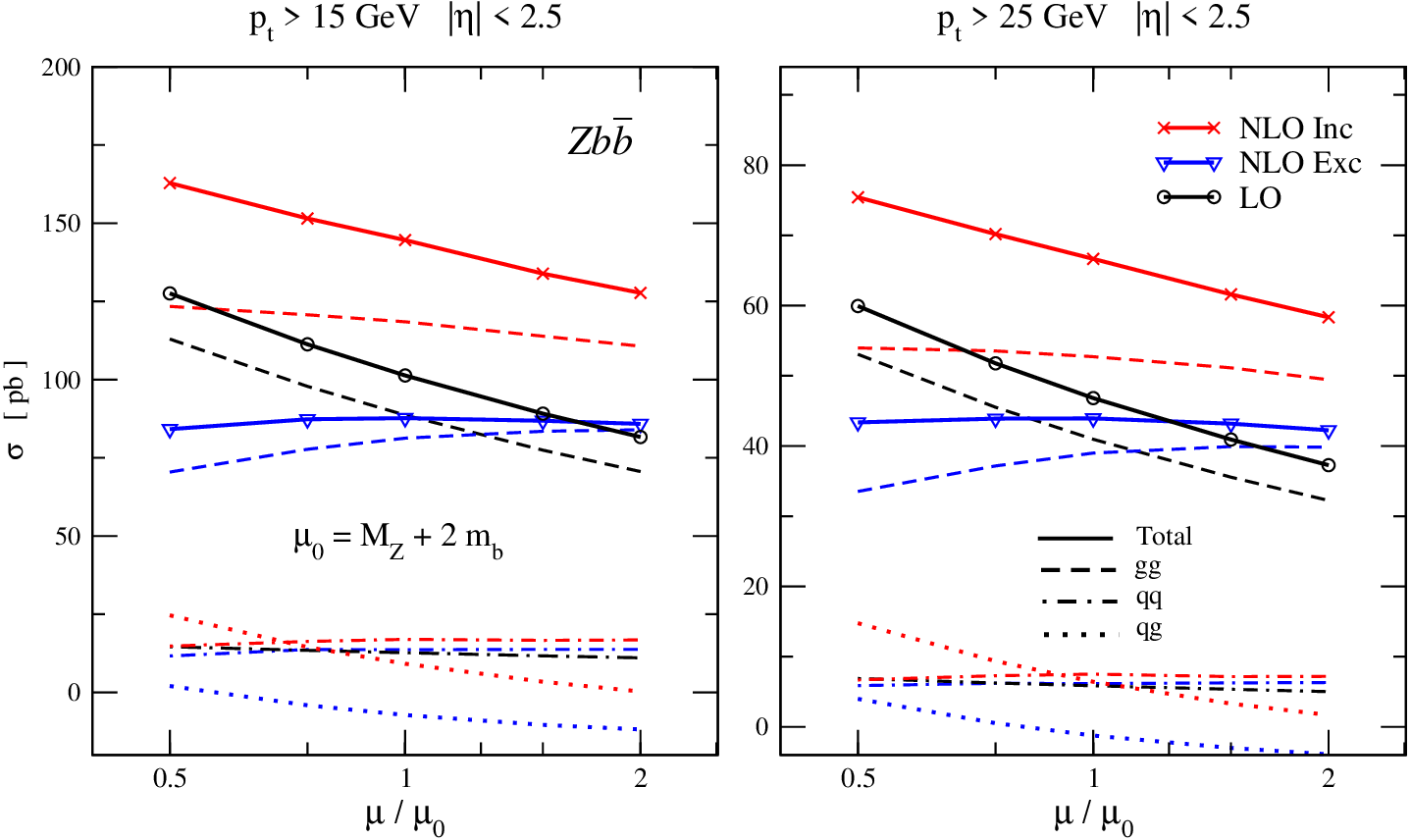} 
\caption[]{Dependence of the LO (black, solid), NLO \emph{exclusive}
  (blue, solid), and NLO \emph{inclusive} (red, solid) total cross
  sections for $Zb\bar{b}$ production on the
  renormalization/factorization scales, including full $b$-quark mass
  effects, when $\mu=\mu_r=\mu_f$ is varied between $\mu_0/2$ and
  $2\mu_0$ (with $\mu_0=M_Z+2m_b$). We also show the individual
  channels, $q\bar{q}^\prime$ (dash-dotted), $qg+\bar{q}g$ (dotted)
  and $gg$ (dashed), for the \emph{inclusive} (red) and
  \emph{exclusive} (blue) cases.}
\label{fig:Zbb_mudep}
\end{center}
\end{figure}

\begin{table}[htb]
\begin{center}
  \caption{LO, NLO \emph{inclusive}, and NLO \emph{exclusive} cross
    sections for $Zb\bar{b}$ at both $\sqrt{s}=10$~TeV and
    $\sqrt{s}=14$~TeV, for two different values of the $b$-jet
    transverse-momentum selection cut, and for zero and non-zero
    $b$-quark mass. The central values correspond to
    $\mu_r=\mu_f=\mu_0=M_Z+2m_b$, while the upper and lower bounds
    represent the maximal upper and lower variations obtained when
    varying $\mu_r=\mu_f$ between $\mu_0/2$ and $2\mu_0$.}
\label{tab:summary_zbb}
\begin{tabular}{|c|c|c|c|c|} \hline\hline
& \multicolumn{2}{|c|}{$p_T^b>15$~GeV} & \multicolumn{2}{|c|}{$p_T^b>25$~GeV}\\ 
& $m_b\neq 0$ & $m_b=0$ & $m_b\neq 0$ & $m_b=0$\\
\hline
&\multicolumn{4}{|c|}{$\sqrt{s}=10$~TeV} \\
\hline
$\sigma_{LO}$ (pb) & $55.1^{+16}_{-12}$ & $57.6^{+18}_{-13}$ & $24.6^{+7.6}_{-5.4}$ & $25.1^{+8.4}_{-5.9}$ \\
$\sigma_{NLO,inc}$ (pb) & $82.5^{+12}_{-11}$ & $84.5^{+14}_{-12}$ & $36.0^{+3.9}_{-4.6}$ & $36.1^{+6.4}_{-5.2}$ \\
$\sigma_{NLO,exc}$ (pb) & $52.1^{+0.0}_{-1.7}$ & $ 53.5^{+0.2}_{-2.4}$ &$24.6^{+0.3}_{-1.2}$ & $24.7^{+0.3}_{-1.6}$\\
\hline
&\multicolumn{4}{|c|}{$\sqrt{s}=14$~TeV} \\
\hline
$\sigma_{LO}$ (pb) & $101^{+26}_{-20}$ & $106^{+30}_{-22}$ & $46.8^{+13.1}_{-9.6}$ & 
$46.8^{+12.7}_{-9.9}$\\
$\sigma_{NLO,inc}$ (pb) & $145^{+20}_{-17}$ & $148^{+24}_{-19}$ & $66.6^{+8.8}_{-8.3}$ & 
$66.1^{+10.5}_{-9.1}$\\
$\sigma_{NLO,exc}$ (pb) & $88.4^{+0.0}_{-3.0}$ & $90.0^{+0.0}_{-1.6}$ & $43.7^{+0.0}_{-1.6}$ & 
$43.5^{+0.4}_{-1.9}$\\
\hline
\end{tabular}
\end{center}
\end{table}

\begin{figure}[htb]
\begin{center}
\includegraphics*[clip,scale=0.5]{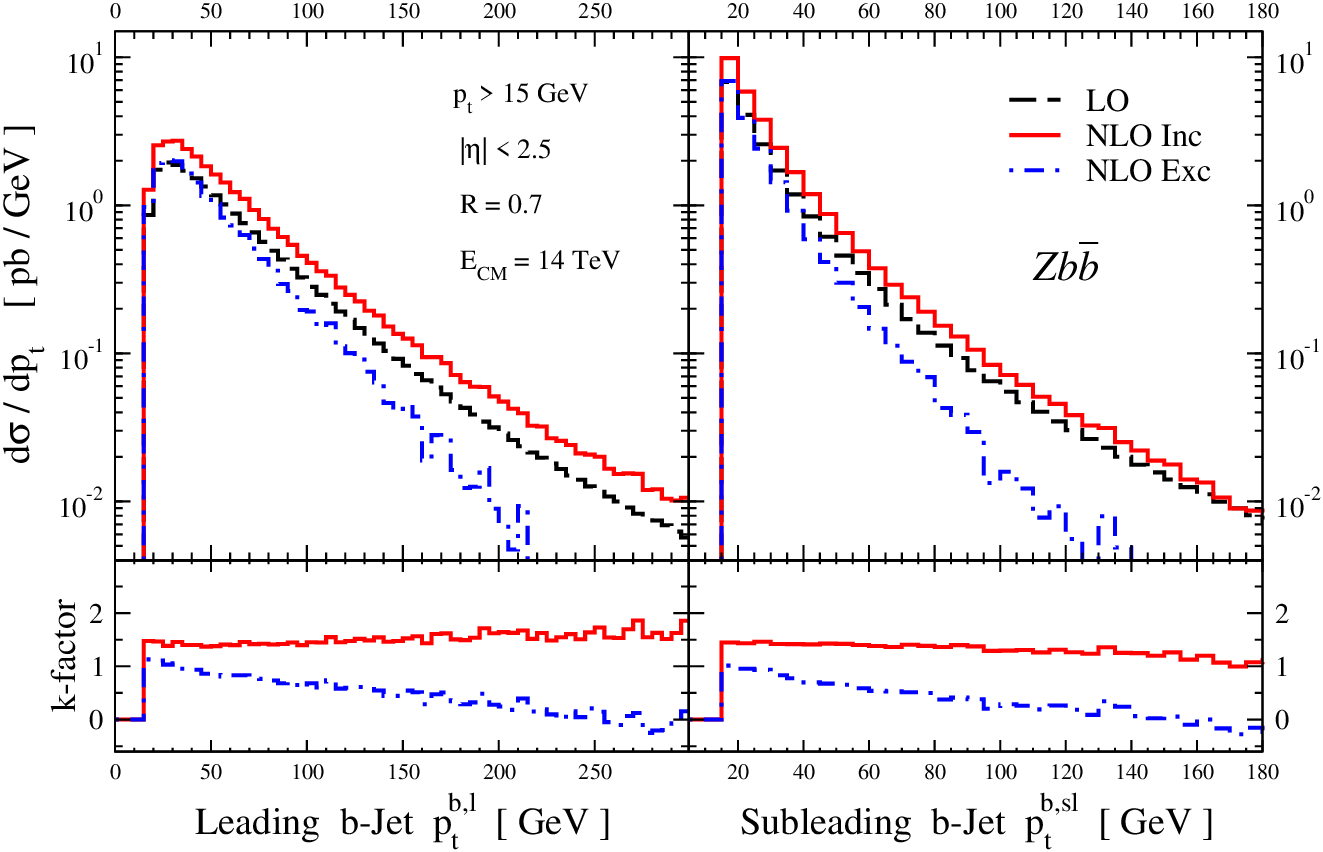} 
\caption[]{LO (black, dashed), NLO \emph{inclusive} (red, solid) and
  NLO \emph{exclusive} (blue, dot-dashed) transverse momentum
  distributions for the $b$ jet with the leading (left hand side)
  and subleading (right hand side) transverse momentum in $Zb\bar{b}$
  production. The lower window shows the bin-by-bin K factor, for the
  \emph{inclusive} (red, solid) and \emph{exclusive} (blue,
  dot-dashed) cases.}
\label{fig:Zbb_pt_bjets_pt15}
\end{center}
\end{figure}
\begin{figure}[htb]
\begin{center}
\includegraphics*[clip,scale=0.46]{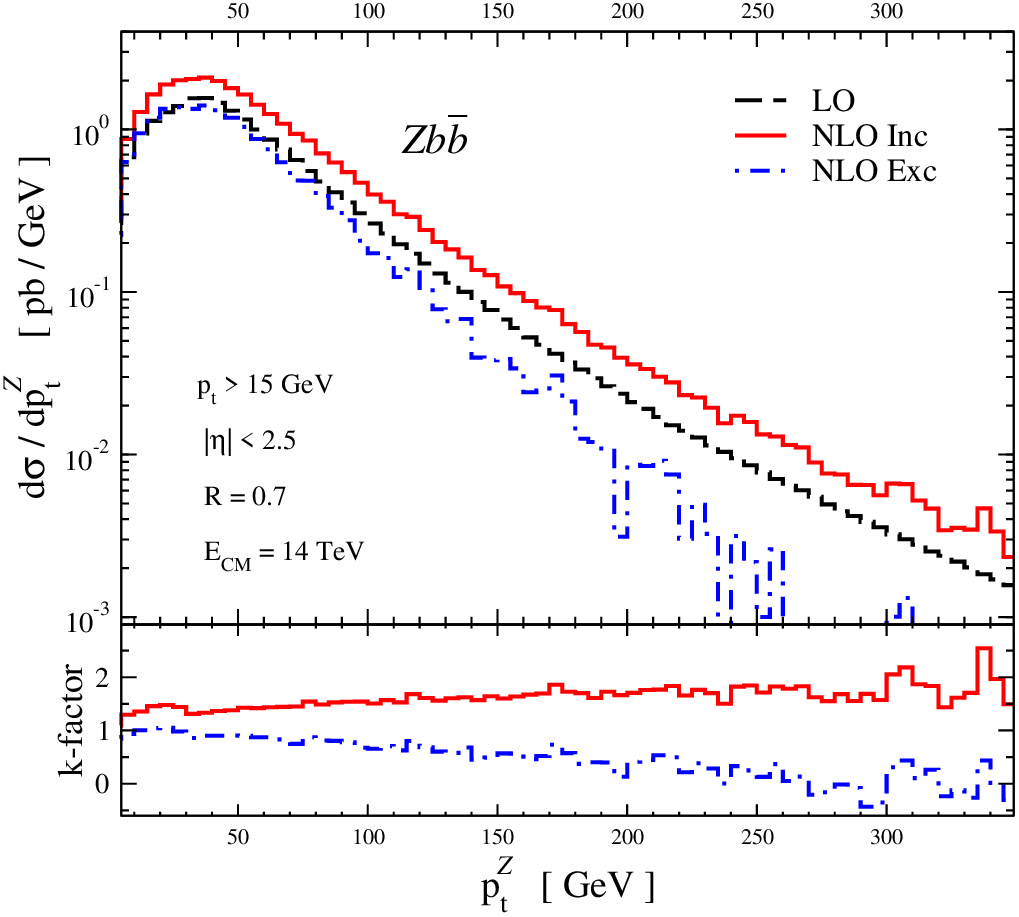} 
\caption[]{LO (black, dashed), NLO \emph{inclusive} (red, solid) and
  NLO \emph{exclusive} (blue, dot-dashed) transverse momentum
  distributions for the $Z$ boson in $Zb\bar{b}$. The lower window
  shows the bin-by-bin K factor, for the \emph{inclusive} (red, solid)
  and \emph{exclusive} (blue, dot-dashed) cases.}
\label{fig:Zbb_pt_Z_pt15}
\end{center}
\end{figure}
\begin{figure}[htb]
\begin{center}
\includegraphics*[clip,scale=0.5]{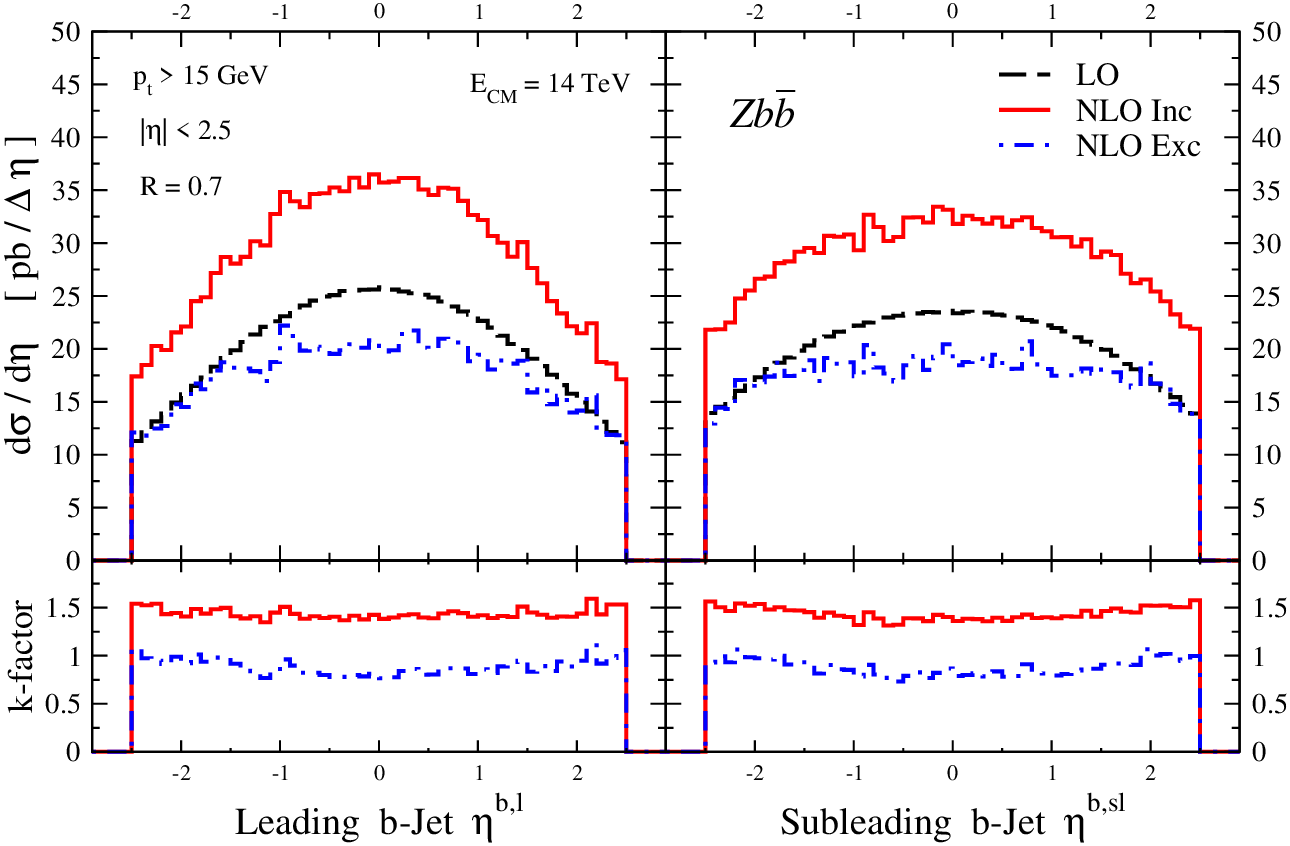}
\caption[]{LO (black, dashed), NLO \emph{inclusive} (red, solid) and
  NLO \emph{exclusive} (blue, dot-dashed) pseudorapidity distributions
  for the $b$ jet with the leading (left hand side) and subleading
  (right hand side) transverse momentum in $Zb\bar{b}$ production. The
  lower window shows the bin-by-bin K factor, for the \emph{inclusive}
  (red, solid) and \emph{exclusive} (blue, dot-dashed) cases.}
\label{fig:Zbb_eta_bjets_pt15}
\end{center}
\end{figure}
\begin{figure}[htb]
\begin{center}
\includegraphics*[clip,scale=0.46]{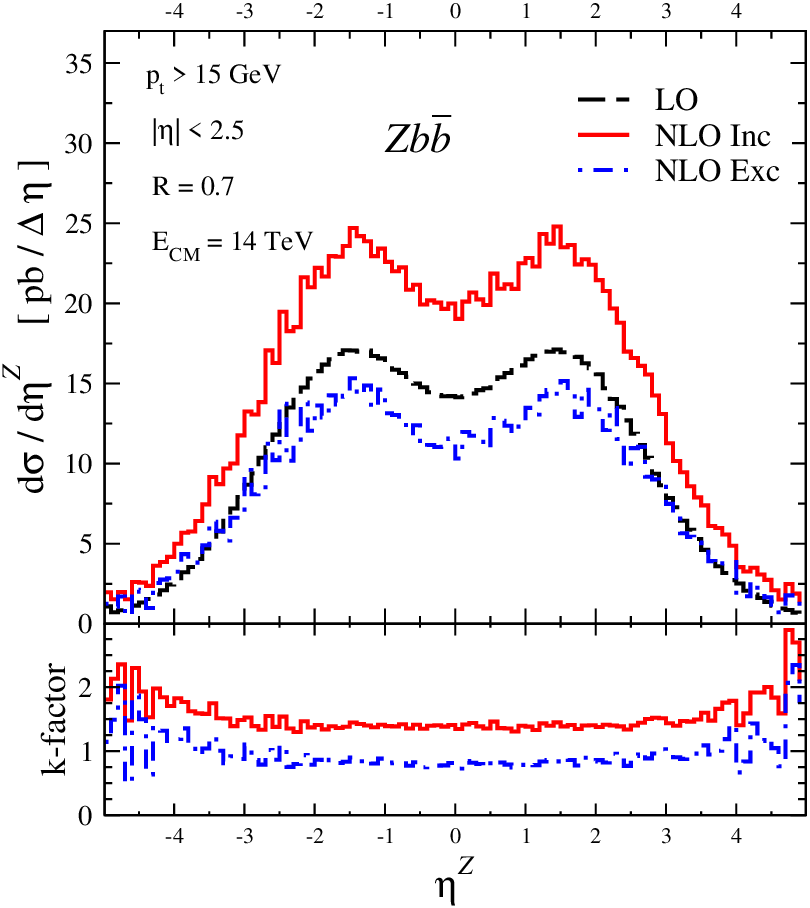} 
\caption[]{LO (black, dashed), NLO \emph{inclusive} (red, solid) and
  NLO \emph{exclusive} (blue, dot-dashed) pseudorapidity distributions
  for the $Z$ boson in $Zb\bar{b}$ production. The lower window shows
  the bin-by-bin K factor, for the \emph{inclusive} (red, solid) and
  \emph{exclusive} (blue, dot-dashed) cases.}
\label{fig:Zbb_eta_Z_pt15}
\end{center}
\end{figure}
\begin{figure}[htb]
\begin{center}
\includegraphics*[clip,scale=0.46]{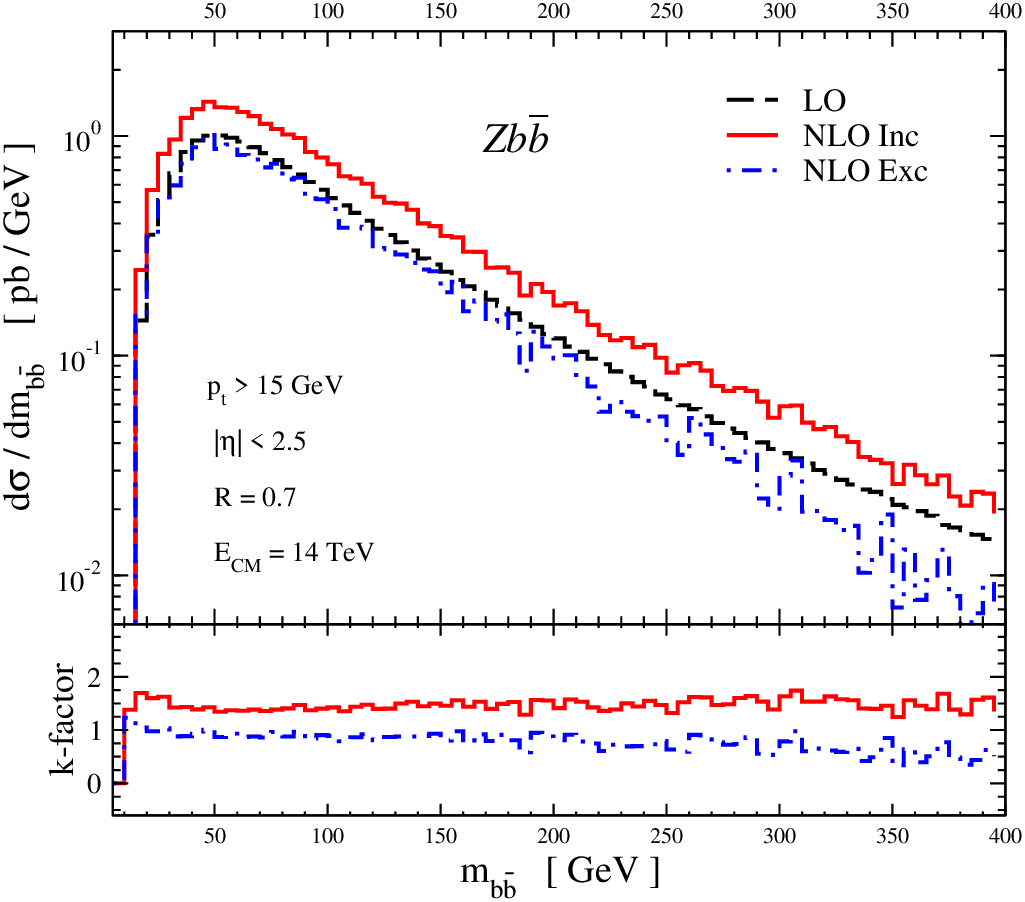} 
\caption[]{LO (black, dashed), NLO \emph{inclusive} (red, solid) and
  NLO \emph{exclusive} (blue, dot-dashed) $b\bar{b}$-pair
  invariant-mass distributions for $Zb\bar{b}$ production. The lower
  window shows the bin-by-bin K factor, for the \emph{inclusive} (red,
  solid) and \emph{exclusive} (blue, dot-dashed) cases.}
\label{fig:Zbb_mbb_pt15}
\end{center}
\end{figure}
\begin{figure}[htb]
\begin{center}
\includegraphics*[clip,scale=0.46]{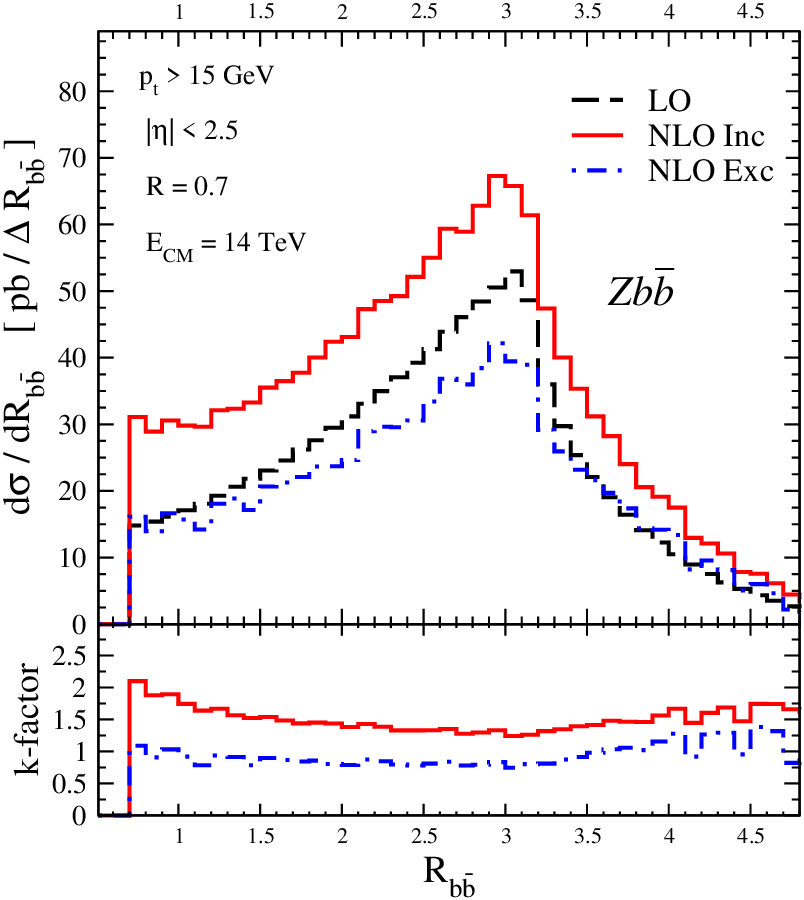} 
\caption[]{LO (black, dashed), NLO \emph{inclusive} (red, solid) and
  NLO \emph{exclusive} (blue, dot-dashed) $b$-jet relative-distance
  distributions for $Zb\bar{b}$ production. The lower window shows the
  bin-by-bin K factor, for the \emph{inclusive} (red, solid) and
  \emph{exclusive} (blue, dot-dashed) cases.}
\label{fig:Zbb_Rbb_pt15}
\end{center}
\end{figure}
\begin{figure}[htb]
\begin{center}
\begin{tabular}{ll}
\includegraphics*[clip,scale=0.52]{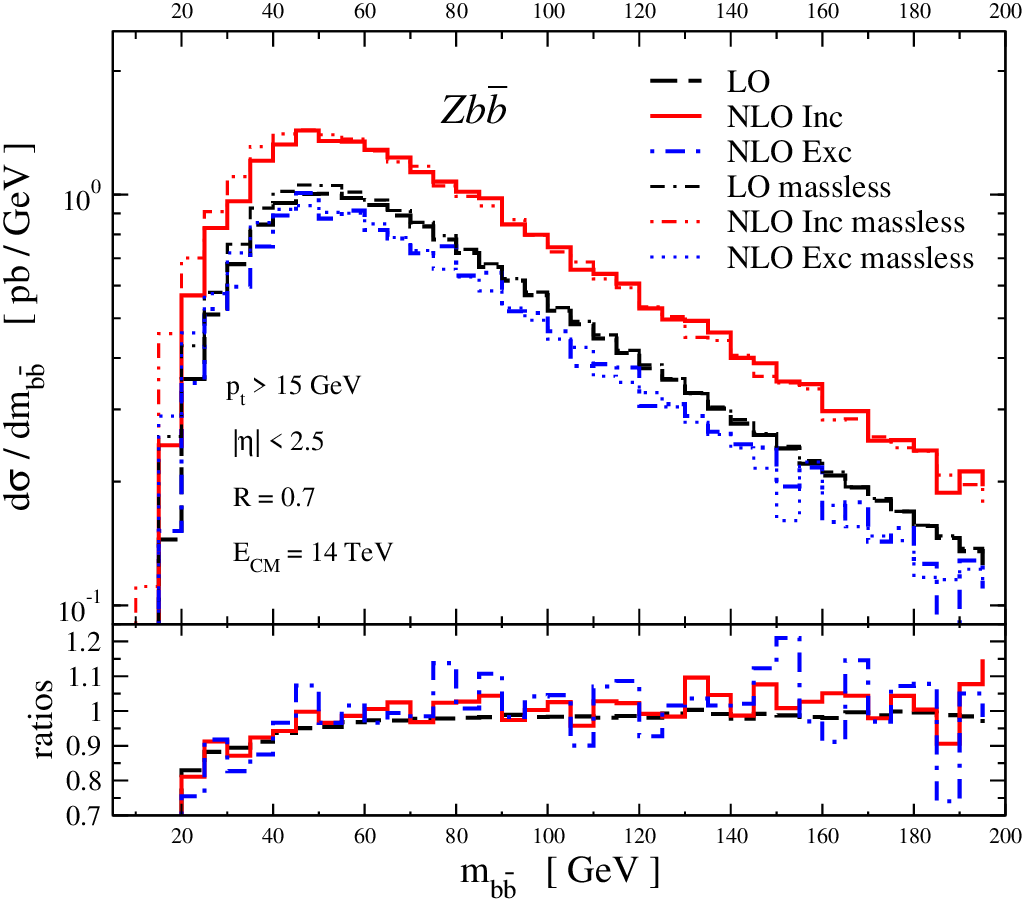} &
\includegraphics*[clip,scale=0.52]{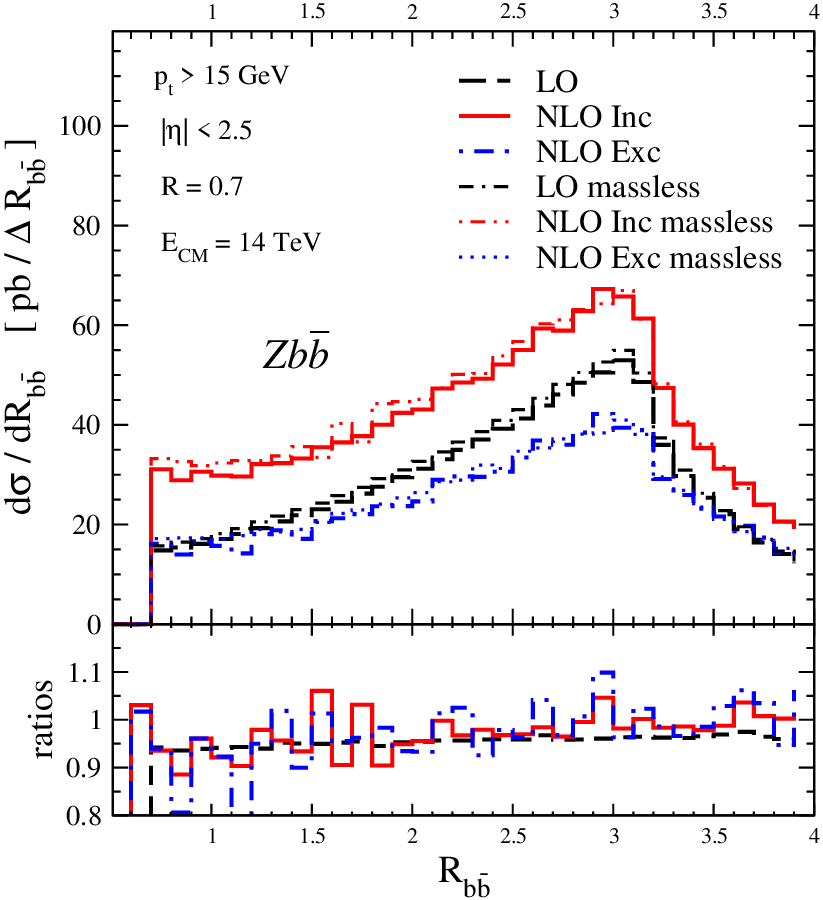} 
\end{tabular} 
\caption[]{LO (black), NLO \emph{inclusive} (red) and NLO
  \emph{exclusive} (blue) $m_{b\bar{b}}$ (left hand side) and
  $R_{b\bar{b}}$ (right hand side) distributions for $Zb\bar{b}$
  production derived from our calculation with $m_b\ne 0$ (LO: dashed,
  NLO \emph{inclusive}: solid, NLO \emph{exclusive}: dash-dotted) and
  from MCFM with $m_b=0$ (LO: double-dashed/dotted, NLO
  \emph{inclusive}: dashed/double-dotted, NLO \emph{exclusive}:
  dotted).  The lower window shows the ratio of the distributions for
  massive and massless $b$ quarks, $d\sigma(m_b\neq 0)/d\sigma(m_b=0)$
  (LO: dashed, NLO \emph{inclusive}: solid, NLO \emph{exclusive}:
  dash-dotted).}
\label{fig:Zbb_comp_pt15}
\end{center}
\end{figure}
\boldmath
\section{$Zb\bar{b}$ production}
\label{sec:zbb}
\unboldmath

At tree level, the production of a $Z$ boson with a pair of bottom
quarks consists of two channels, namely $q\bar{q}\rightarrow
Zb\bar{b}$ and $gg\rightarrow Zb\bar{b}$. At NLO in QCD one needs to
include the one-loop virtual corrections to both tree-level processes
as well as the real radiation corrections with up to one extra parton
in the final state, i.e.  $q\bar{q}\rightarrow Zb\bar{b}+g$,
$gg\rightarrow Zb\bar{b}+g$, and $qg(\bar{q}g)\rightarrow
Zb\bar{b}+q(\bar{q})$.  Details of the calculation have been given in
Refs.~\cite{FebresCordero:2008ci,Cordero:2008ce} and will not be
repeated here.

As done in the case of $W^+b\bar{b}$ and $W^-b\bar{b}$ production in
Section~\ref{sec:wbb}, we start our discussion of $Zb\bar b$ cross
sections at NLO QCD at the LHC by studying the renormalization- and
factorization-scale dependence of the LO and NLO total cross sections.
In Fig.~\ref{fig:Zbb_mudep} we show the scale dependence of the
individual parton-level channels, as well as their sum, at LO and NLO
in QCD for both the \emph{inclusive} and \emph{exclusive} case.  As
for $Wb\bar{b}$ production, we notice the pronounced scale
dependence of the NLO total cross section to $qg(\bar{q}g)\rightarrow
Zb\bar{b}+q(\bar{q})$.  However, since in $Zb\bar{b}$ production both
the LO and NLO cross sections also consist of a $gg$-initiated
subprocess, the $qg(\bar{q}g)\rightarrow Zb\bar{b}+q(\bar{q})$
channel is not dominant at NLO and its effect is therefore less
pronounced. Indeed, the scale dependence of the \emph{exclusive} cross
section actually greatly improves at NLO in QCD, while the scale
dependence of the \emph{inclusive} one is only mildly better than at
LO, but not worse as it is the case in $Wb\bar b$ production.

In Table~\ref{tab:summary_zbb} we provide both LO and NLO total cross
sections for $Zb\bar{b}$ production, including our estimate of the
residual uncertainty due to only the renormalization- and
factorization-scale dependence. We give results for two choices of
$p_T^b$ cuts, $p_T^b>15$~GeV and $p_T^b>25$~GeV, and for both
center-of-mass energies, $\sqrt{s}=14$~TeV and $\sqrt{s}=10$~TeV.  We
compare the case of a massless and massive bottom quark and we observe
that the effects of a non-zero $b$-quark mass are mild for $Zb\bar b$
production.  Indeed, for $p_t^b>15$~GeV the massless approximation
mildly overestimates the total cross section by about $2-3\%$, which in
the \emph{inclusive} case is considerably smaller than the scale uncertainty.
For $p_t^b>25$~GeV the difference is basically gone. We remind,
however, that this is not the case in more inclusive studies, like
$Z+1\,b$-jet production~\cite{Zb_inprep}, where, as seen in
$W+1\,b$-jet production~\cite{Campbell:2008hh}, one needs to
consistently include full $b$-quark mass effects.

In Figs.~(\ref{fig:Zbb_pt_bjets_pt15})-(\ref{fig:Zbb_eta_Z_pt15}) we
show the transverse momentum ($p_T$) and pseudorapidity ($\eta$)
distributions for the $b$ jet with the leading and subleading
transverse momentum, and for the $Z$ boson. As in the $Wb\bar b$ case,
the upper parts of each figure show the LO, NLO \emph{inclusive} and
NLO \emph{exclusive} distributions, while in the lower parts we
provide the bin-by-bin K factors. We show results obtained by using
the central scale choice $\mu_r=\mu_f=\mu_0=M_Z+2m_b$.  As expected
from the study of the scale dependence shown in
Fig.~(\ref{fig:Zbb_mudep}), in the \emph{exclusive} case the NLO QCD
corrections reduce the cross sections as compared to the LO result,
while in the \emph{inclusive} case the NLO QCD corrections increase
them.  Figures~(\ref{fig:Zbb_pt_bjets_pt15}) and (\ref{fig:Zbb_pt_Z_pt15})
also show, in particular in the bin-by-bin K factors, that the NLO QCD
corrections affect the shape of the \emph{exclusive} transverse
momentum distributions, with large negative corrections at large $p_T$ where
the bin-by-bin K factors become smaller than one, while
the changes in the shape are much less pronounced in the \emph{inclusive} case.
Figures~(\ref{fig:Zbb_eta_bjets_pt15}) and (\ref{fig:Zbb_eta_Z_pt15})
show that NLO QCD corrections have hardly any effects on the shape of
the pseudorapidity distributions. The distributions for the invariant mass
of the two $b$ jets ($m_{b\bar{b}}$) and for their relative distance
($R_{b\bar{b}}$) are shown in Figs.~(\ref{fig:Zbb_mbb_pt15}) and
(\ref{fig:Zbb_Rbb_pt15}) respectively. We notice that, although in a
much milder fashion than for $Wb\bar{b}$ production, in the
\emph{inclusive} case NLO QCD corrections enhance the regions at small
and large values of $R_{b\bar{b}}$.

Finally, in Fig.~\ref{fig:Zbb_comp_pt15} we compare the LO and NLO
$m_{b\bar{b}}$ and $R_{b\bar{b}}$ distributions for $Z b\bar b$
production obtained from the massive and massless $b$-quark
calculations. The results with $m_b=0$ have been obtained using MCFM.
Most of the difference is seen in the region of low invariant mass
$m_{b\bar{b}}$, and are more noticeable for small values of
$R_{b\bar{b}}$.  In all regions though the impact of a non-zero
$b$-quark mass is almost not affected by including NLO QCD
corrections, and can be taken into account by rescaling the NLO result
for massless bottom quarks with the ratio of the LO cross sections as
discussed in~\cite{FebresCordero:2008ci}.

\section{Conclusions}
\label{sec:conclusions}

A reliable theoretical prediction for $W$ and $Z$ production with $b$
jets will be crucial for many Higgs-boson searches and studies at the
LHC.  In this paper we have presented a study of $W$ and $Z$ boson
production with two $b$ jets at the LHC including full $b$-quark mass
effects, based on the NLO QCD calculation of $Wb\bar{b}$ and
$Zb\bar{b}$ production presented in
Refs.~\cite{FebresCordero:2006sj,FebresCordero:2008ci,Cordero:2008ce}. We
have presented numerical results for the total $Wb\bar b$ and $Zb \bar
b$ production cross sections, as well as for a number of kinematic
distributions interesting to LHC physics, for both massive (our
calculation) and massless bottom quarks (as implemented in MCFM).  The
$Wb\bar{b}$ cross sections at NLO QCD still suffer from a large
theoretical uncertainty due to the unphysical renormalization- and
factorization-scale dependence, which is particularly pronounced in
the \emph{inclusive} case, and further theoretical improvements are
warranted.  In the case of $Zb\bar{b}$ production, the NLO QCD cross
sections are well behaved, i.~e. exhibit only a mild residual scale
dependence, in particular in the \emph{exclusive} case.  The shape of
distributions is changed significantly by NLO QCD corrections, such
that they can not be correctly described by global K-factor
rescalings.  This is more dramatic in the case for $Wb\bar{b}$
production, in view of which including resummation effects, as well as
consistent NLO showering, might be of considerable importance.

The $b$-quark mass effects can impact the shape of the kinematic
distributions, as shown on the example of the $m_{b\bar b}$ and
$R_{b\bar b}$ distributions, in particular in phase space regions
where the relevant kinematic observable is of the order of
$m_b$. Apart from these regions, however, these effects can be
approximated by rescaling the NLO cross section for $m_b=0$ with the
ratio of LO cross sections for massive and massless bottom quarks as
discussed in detail
in~\cite{FebresCordero:2006sj,FebresCordero:2008ci}. The total
production cross sections are reduced by $b$-quark mass effects, and
the effect is more pronounced the smaller the applied $p_T^b$
cut. However, these effects are in most cases smaller than the
residual scale dependence at NLO in QCD, especially in $Wb\bar b$
production for the \emph{inclusive} case.

\section*{Acknowledgements}

F.~F.~C. and L.~R. would like to thank the Theory Division of CERN for
its kind hospitality and support during the \emph{From LHC to Future
  Colliders} CERN Theory Institute, where part of the work presented
in this paper was completed. We thank in particular Chiara Mariotti,
Sacha Nikitenko and the CMS Higgs working group for their interest and
for very useful discussions. F.~F.~C. thanks Harald Ita for helpful
discussions.  The work of F.~F.~C. and L.~R.~is supported in part by
the U.S. Department of Energy under grants DE-FG03-91ER40662 and
DE-FG02-97IR41022 respectively.  The work of D.~W.~is supported in
part by the National Science Foundation under grants NSF-PHY-0757691
and NSF-PHY-0547564.

\bibliography{wzbb_lhc}

\begin{thebibliography}{50}
\expandafter\ifx\csname natexlab\endcsname\relax\def\natexlab#1{#1}\fi
\expandafter\ifx\csname bibnamefont\endcsname\relax
  \def\bibnamefont#1{#1}\fi
\expandafter\ifx\csname bibfnamefont\endcsname\relax
  \def\bibfnamefont#1{#1}\fi
\expandafter\ifx\csname citenamefont\endcsname\relax
  \def\citenamefont#1{#1}\fi
\expandafter\ifx\csname url\endcsname\relax
  \def\url#1{\texttt{#1}}\fi
\expandafter\ifx\csname urlprefix\endcsname\relax\def\urlprefix{URL }\fi
\providecommand{\bibinfo}[2]{#2}
\providecommand{\eprint}[2][]{\url{#2}}

\bibitem[{\citenamefont{Aaltonen et~al.}(2008)}]{Aaltonen:2008mt}
\bibinfo{author}{\bibfnamefont{T.}~\bibnamefont{Aaltonen}} \bibnamefont{et~al.}
  (\bibinfo{collaboration}{CDF}) (\bibinfo{year}{2008}),
  \eprint{arXiv:0812.4458 [hep-ex]}.

\bibitem[{\citenamefont{Neu et~al.}(2008)\citenamefont{Neu, Thomson, and
  Heinrich}}]{D0Wbb:2008}
\bibinfo{author}{\bibfnamefont{C.}~\bibnamefont{Neu}},
  \bibinfo{author}{\bibfnamefont{E.}~\bibnamefont{Thomson}}, \bibnamefont{and}
  \bibinfo{author}{\bibfnamefont{J.}~\bibnamefont{Heinrich}}
  (\bibinfo{collaboration}{CDF}) (\bibinfo{year}{2008}), \eprint{CDF note
  9321}.

\bibitem[{\citenamefont{Abazov et~al.}(2005{\natexlab{a}})}]{Abazov:2004jy}
\bibinfo{author}{\bibfnamefont{V.~M.} \bibnamefont{Abazov}}
  \bibnamefont{et~al.} (\bibinfo{collaboration}{D0}), \bibinfo{journal}{Phys.
  Rev. Lett.} \textbf{\bibinfo{volume}{94}}, \bibinfo{pages}{091802}
  (\bibinfo{year}{2005}{\natexlab{a}}), \eprint{hep-ex/0410062}.

\bibitem[{\citenamefont{Abazov et~al.}(2005{\natexlab{b}})}]{Abazov:2004zd}
\bibinfo{author}{\bibfnamefont{V.~M.} \bibnamefont{Abazov}}
  \bibnamefont{et~al.} (\bibinfo{collaboration}{D0}), \bibinfo{journal}{Phys.
  Rev. Lett.} \textbf{\bibinfo{volume}{94}}, \bibinfo{pages}{161801}
  (\bibinfo{year}{2005}{\natexlab{b}}), \eprint{hep-ex/0410078}.

\bibitem[{\citenamefont{Kao et~al.}(2007)\citenamefont{Kao, Dicus, Malhotra,
  and Wang}}]{Kao:2007qw}
\bibinfo{author}{\bibfnamefont{C.}~\bibnamefont{Kao}},
  \bibinfo{author}{\bibfnamefont{D.~A.} \bibnamefont{Dicus}},
  \bibinfo{author}{\bibfnamefont{R.}~\bibnamefont{Malhotra}}, \bibnamefont{and}
  \bibinfo{author}{\bibfnamefont{Y.}~\bibnamefont{Wang}}
  (\bibinfo{year}{2007}), \eprint{arXiv:0711.0232 [hep-ph]}.

\bibitem[{\citenamefont{Adam et~al.}(2008)}]{:2008uu}
\bibinfo{author}{\bibfnamefont{N.~E.} \bibnamefont{Adam}} \bibnamefont{et~al.}
  (\bibinfo{year}{2008}), \eprint{arXiV:0803.1154 [hep-ph]}.

\bibitem[{\citenamefont{Harlander}(2000)}]{Harlander:2000mg}
\bibinfo{author}{\bibfnamefont{R.~V.} \bibnamefont{Harlander}},
  \bibinfo{journal}{Phys. Lett.} \textbf{\bibinfo{volume}{B492}},
  \bibinfo{pages}{74} (\bibinfo{year}{2000}), \eprint{hep-ph/0007289}.

\bibitem[{\citenamefont{Harlander and Kilgore}(2001)}]{Harlander:2001is}
\bibinfo{author}{\bibfnamefont{R.~V.} \bibnamefont{Harlander}}
  \bibnamefont{and} \bibinfo{author}{\bibfnamefont{W.~B.}
  \bibnamefont{Kilgore}}, \bibinfo{journal}{Phys. Rev.}
  \textbf{\bibinfo{volume}{D64}}, \bibinfo{pages}{013015}
  (\bibinfo{year}{2001}), \eprint{hep-ph/0102241}.

\bibitem[{\citenamefont{Harlander and Kilgore}(2002)}]{Harlander:2002wh}
\bibinfo{author}{\bibfnamefont{R.~V.} \bibnamefont{Harlander}}
  \bibnamefont{and} \bibinfo{author}{\bibfnamefont{W.~B.}
  \bibnamefont{Kilgore}}, \bibinfo{journal}{Phys. Rev. Lett.}
  \textbf{\bibinfo{volume}{88}}, \bibinfo{pages}{201801}
  (\bibinfo{year}{2002}), \eprint{hep-ph/0201206}.

\bibitem[{\citenamefont{Anastasiou and Melnikov}(2002)}]{Anastasiou:2002yz}
\bibinfo{author}{\bibfnamefont{C.}~\bibnamefont{Anastasiou}} \bibnamefont{and}
  \bibinfo{author}{\bibfnamefont{K.}~\bibnamefont{Melnikov}},
  \bibinfo{journal}{Nucl. Phys.} \textbf{\bibinfo{volume}{B646}},
  \bibinfo{pages}{220} (\bibinfo{year}{2002}), \eprint{hep-ph/0207004}.

\bibitem[{\citenamefont{Anastasiou et~al.}(2004)\citenamefont{Anastasiou,
  Melnikov, and Petriello}}]{Anastasiou:2004xq}
\bibinfo{author}{\bibfnamefont{C.}~\bibnamefont{Anastasiou}},
  \bibinfo{author}{\bibfnamefont{K.}~\bibnamefont{Melnikov}}, \bibnamefont{and}
  \bibinfo{author}{\bibfnamefont{F.}~\bibnamefont{Petriello}},
  \bibinfo{journal}{Phys. Rev. Lett.} \textbf{\bibinfo{volume}{93}},
  \bibinfo{pages}{262002} (\bibinfo{year}{2004}), \eprint{hep-ph/0409088}.

\bibitem[{\citenamefont{Ravindran et~al.}(2003)\citenamefont{Ravindran, Smith,
  and van Neerven}}]{Ravindran:2003um}
\bibinfo{author}{\bibfnamefont{V.}~\bibnamefont{Ravindran}},
  \bibinfo{author}{\bibfnamefont{J.}~\bibnamefont{Smith}}, \bibnamefont{and}
  \bibinfo{author}{\bibfnamefont{W.~L.} \bibnamefont{van Neerven}},
  \bibinfo{journal}{Nucl. Phys.} \textbf{\bibinfo{volume}{B665}},
  \bibinfo{pages}{325} (\bibinfo{year}{2003}), \eprint{hep-ph/0302135}.

\bibitem[{\citenamefont{Catani et~al.}(2001)\citenamefont{Catani, de~Florian,
  and Grazzini}}]{Catani:2001ic}
\bibinfo{author}{\bibfnamefont{S.}~\bibnamefont{Catani}},
  \bibinfo{author}{\bibfnamefont{D.}~\bibnamefont{de~Florian}},
  \bibnamefont{and} \bibinfo{author}{\bibfnamefont{M.}~\bibnamefont{Grazzini}},
  \bibinfo{journal}{JHEP} \textbf{\bibinfo{volume}{05}}, \bibinfo{pages}{025}
  (\bibinfo{year}{2001}), \eprint{hep-ph/0102227}.

\bibitem[{\citenamefont{Catani et~al.}(2003)\citenamefont{Catani, de~Florian,
  Grazzini, and Nason}}]{Catani:2003zt}
\bibinfo{author}{\bibfnamefont{S.}~\bibnamefont{Catani}},
  \bibinfo{author}{\bibfnamefont{D.}~\bibnamefont{de~Florian}},
  \bibinfo{author}{\bibfnamefont{M.}~\bibnamefont{Grazzini}}, \bibnamefont{and}
  \bibinfo{author}{\bibfnamefont{P.}~\bibnamefont{Nason}},
  \bibinfo{journal}{JHEP} \textbf{\bibinfo{volume}{07}}, \bibinfo{pages}{028}
  (\bibinfo{year}{2003}), \eprint{hep-ph/0306211}.

\bibitem[{\citenamefont{Bozzi et~al.}(2006)\citenamefont{Bozzi, Catani,
  de~Florian, and Grazzini}}]{Bozzi:2005wk}
\bibinfo{author}{\bibfnamefont{G.}~\bibnamefont{Bozzi}},
  \bibinfo{author}{\bibfnamefont{S.}~\bibnamefont{Catani}},
  \bibinfo{author}{\bibfnamefont{D.}~\bibnamefont{de~Florian}},
  \bibnamefont{and} \bibinfo{author}{\bibfnamefont{M.}~\bibnamefont{Grazzini}},
  \bibinfo{journal}{Nucl. Phys.} \textbf{\bibinfo{volume}{B737}},
  \bibinfo{pages}{73} (\bibinfo{year}{2006}), \eprint{hep-ph/0508068}.

\bibitem[{\citenamefont{Anastasiou et~al.}(2009)\citenamefont{Anastasiou,
  Boughezal, and Petriello}}]{Anastasiou:2008tj}
\bibinfo{author}{\bibfnamefont{C.}~\bibnamefont{Anastasiou}},
  \bibinfo{author}{\bibfnamefont{R.}~\bibnamefont{Boughezal}},
  \bibnamefont{and}
  \bibinfo{author}{\bibfnamefont{F.}~\bibnamefont{Petriello}},
  \bibinfo{journal}{JHEP} \textbf{\bibinfo{volume}{04}}, \bibinfo{pages}{003}
  (\bibinfo{year}{2009}), \eprint{arXiv:0811.3458 [hep-ph]}.

\bibitem[{\citenamefont{de~Florian and Grazzini}(2009)}]{deFlorian:2009hc}
\bibinfo{author}{\bibfnamefont{D.}~\bibnamefont{de~Florian}} \bibnamefont{and}
  \bibinfo{author}{\bibfnamefont{M.}~\bibnamefont{Grazzini}},
  \bibinfo{journal}{Phys. Lett.} \textbf{\bibinfo{volume}{B674}},
  \bibinfo{pages}{291} (\bibinfo{year}{2009}), \eprint{arXiv:0901.2427
  [hep-ph]}.

\bibitem[{\citenamefont{Han and Willenbrock}(1991)}]{Han:1991ia}
\bibinfo{author}{\bibfnamefont{T.}~\bibnamefont{Han}} \bibnamefont{and}
  \bibinfo{author}{\bibfnamefont{S.}~\bibnamefont{Willenbrock}},
  \bibinfo{journal}{Phys. Lett.} \textbf{\bibinfo{volume}{B273}},
  \bibinfo{pages}{167} (\bibinfo{year}{1991}).

\bibitem[{\citenamefont{Mrenna and Yuan}(1998)}]{Mrenna:1997wp}
\bibinfo{author}{\bibfnamefont{S.}~\bibnamefont{Mrenna}} \bibnamefont{and}
  \bibinfo{author}{\bibfnamefont{C.~P.} \bibnamefont{Yuan}},
  \bibinfo{journal}{Phys. Lett.} \textbf{\bibinfo{volume}{B416}},
  \bibinfo{pages}{200} (\bibinfo{year}{1998}), \eprint{hep-ph/9703224}.

\bibitem[{\citenamefont{Brein et~al.}(2004)\citenamefont{Brein, Djouadi, and
  Harlander}}]{Brein:2003wg}
\bibinfo{author}{\bibfnamefont{O.}~\bibnamefont{Brein}},
  \bibinfo{author}{\bibfnamefont{A.}~\bibnamefont{Djouadi}}, \bibnamefont{and}
  \bibinfo{author}{\bibfnamefont{R.}~\bibnamefont{Harlander}},
  \bibinfo{journal}{Phys. Lett.} \textbf{\bibinfo{volume}{B579}},
  \bibinfo{pages}{149} (\bibinfo{year}{2004}), \eprint{hep-ph/0307206}.

\bibitem[{\citenamefont{Actis et~al.}(2008)\citenamefont{Actis, Passarino,
  Sturm, and Uccirati}}]{Actis:2008ug}
\bibinfo{author}{\bibfnamefont{S.}~\bibnamefont{Actis}},
  \bibinfo{author}{\bibfnamefont{G.}~\bibnamefont{Passarino}},
  \bibinfo{author}{\bibfnamefont{C.}~\bibnamefont{Sturm}}, \bibnamefont{and}
  \bibinfo{author}{\bibfnamefont{S.}~\bibnamefont{Uccirati}},
  \bibinfo{journal}{Phys. Lett.} \textbf{\bibinfo{volume}{B670}},
  \bibinfo{pages}{12} (\bibinfo{year}{2008}), \eprint{arXiv:0809.1301
  [hep-ph]}.

\bibitem[{\citenamefont{Ciccolini et~al.}(2003)\citenamefont{Ciccolini,
  Dittmaier, and Kr{\"a}mer}}]{Ciccolini:2003jy}
\bibinfo{author}{\bibfnamefont{M.~L.} \bibnamefont{Ciccolini}},
  \bibinfo{author}{\bibfnamefont{S.}~\bibnamefont{Dittmaier}},
  \bibnamefont{and}
  \bibinfo{author}{\bibfnamefont{M.}~\bibnamefont{Kr{\"a}mer}},
  \bibinfo{journal}{Phys. Rev.} \textbf{\bibinfo{volume}{D68}},
  \bibinfo{pages}{073003} (\bibinfo{year}{2003}), \eprint{hep-ph/0306234}.

\bibitem[{\citenamefont{Dawson et~al.}(2004)\citenamefont{Dawson, Jackson,
  Reina, and Wackeroth}}]{Dawson:2003kb}
\bibinfo{author}{\bibfnamefont{S.}~\bibnamefont{Dawson}},
  \bibinfo{author}{\bibfnamefont{C.~B.} \bibnamefont{Jackson}},
  \bibinfo{author}{\bibfnamefont{L.}~\bibnamefont{Reina}}, \bibnamefont{and}
  \bibinfo{author}{\bibfnamefont{D.}~\bibnamefont{Wackeroth}},
  \bibinfo{journal}{Phys. Rev.} \textbf{\bibinfo{volume}{D69}},
  \bibinfo{pages}{074027} (\bibinfo{year}{2004}), \eprint{hep-ph/0311067}.

\bibitem[{\citenamefont{Dittmaier et~al.}(2004)\citenamefont{Dittmaier,
  Kr{\"a}mer, and Spira}}]{Dittmaier:2003ej}
\bibinfo{author}{\bibfnamefont{S.}~\bibnamefont{Dittmaier}},
  \bibinfo{author}{\bibfnamefont{M.}~\bibnamefont{Kr{\"a}mer}},
  \bibnamefont{and} \bibinfo{author}{\bibfnamefont{M.}~\bibnamefont{Spira}},
  \bibinfo{journal}{Phys. Rev.} \textbf{\bibinfo{volume}{D70}},
  \bibinfo{pages}{074010} (\bibinfo{year}{2004}), \eprint{hep-ph/0309204}.

\bibitem[{\citenamefont{Dawson et~al.}(2005{\natexlab{a}})\citenamefont{Dawson,
  Jackson, Reina, and Wackeroth}}]{Dawson:2004sh}
\bibinfo{author}{\bibfnamefont{S.}~\bibnamefont{Dawson}},
  \bibinfo{author}{\bibfnamefont{C.~B.} \bibnamefont{Jackson}},
  \bibinfo{author}{\bibfnamefont{L.}~\bibnamefont{Reina}}, \bibnamefont{and}
  \bibinfo{author}{\bibfnamefont{D.}~\bibnamefont{Wackeroth}},
  \bibinfo{journal}{Phys. Rev. Lett.} \textbf{\bibinfo{volume}{94}},
  \bibinfo{pages}{031802} (\bibinfo{year}{2005}{\natexlab{a}}),
  \eprint{hep-ph/0408077}.

\bibitem[{\citenamefont{Dawson et~al.}(2005{\natexlab{b}})\citenamefont{Dawson,
  Jackson, Reina, and Wackeroth}}]{Dawson:2004wq}
\bibinfo{author}{\bibfnamefont{S.}~\bibnamefont{Dawson}},
  \bibinfo{author}{\bibfnamefont{C.~B.} \bibnamefont{Jackson}},
  \bibinfo{author}{\bibfnamefont{L.}~\bibnamefont{Reina}}, \bibnamefont{and}
  \bibinfo{author}{\bibfnamefont{D.}~\bibnamefont{Wackeroth}},
  \bibinfo{journal}{Int. J. Mod. Phys.} \textbf{\bibinfo{volume}{A20}},
  \bibinfo{pages}{3353} (\bibinfo{year}{2005}{\natexlab{b}}),
  \eprint{hep-ph/0409345}.

\bibitem[{\citenamefont{Campbell
  et~al.}(2004{\natexlab{a}})\citenamefont{Campbell, Ellis, Maltoni, and
  Willenbrock}}]{Campbell:2003dd}
\bibinfo{author}{\bibfnamefont{J.~M.} \bibnamefont{Campbell}},
  \bibinfo{author}{\bibfnamefont{R.~K.} \bibnamefont{Ellis}},
  \bibinfo{author}{\bibfnamefont{F.}~\bibnamefont{Maltoni}}, \bibnamefont{and}
  \bibinfo{author}{\bibfnamefont{S.}~\bibnamefont{Willenbrock}},
  \bibinfo{journal}{Phys. Rev.} \textbf{\bibinfo{volume}{D69}},
  \bibinfo{pages}{074021} (\bibinfo{year}{2004}{\natexlab{a}}),
  \eprint{hep-ph/0312024}.

\bibitem[{\citenamefont{Campbell et~al.}(2006)\citenamefont{Campbell, Ellis,
  Maltoni, and Willenbrock}}]{Campbell:2005zv}
\bibinfo{author}{\bibfnamefont{J.}~\bibnamefont{Campbell}},
  \bibinfo{author}{\bibfnamefont{R.~K.} \bibnamefont{Ellis}},
  \bibinfo{author}{\bibfnamefont{F.}~\bibnamefont{Maltoni}}, \bibnamefont{and}
  \bibinfo{author}{\bibfnamefont{S.}~\bibnamefont{Willenbrock}},
  \bibinfo{journal}{Phys. Rev.} \textbf{\bibinfo{volume}{D73}},
  \bibinfo{pages}{054007} (\bibinfo{year}{2006}), \eprint{hep-ph/0510362}.

\bibitem[{\citenamefont{Campbell et~al.}(2007)\citenamefont{Campbell, Ellis,
  Maltoni, and Willenbrock}}]{Campbell:2006cu}
\bibinfo{author}{\bibfnamefont{J.}~\bibnamefont{Campbell}},
  \bibinfo{author}{\bibfnamefont{R.~K.} \bibnamefont{Ellis}},
  \bibinfo{author}{\bibfnamefont{F.}~\bibnamefont{Maltoni}}, \bibnamefont{and}
  \bibinfo{author}{\bibfnamefont{S.}~\bibnamefont{Willenbrock}},
  \bibinfo{journal}{Phys. Rev.} \textbf{\bibinfo{volume}{D75}},
  \bibinfo{pages}{054015} (\bibinfo{year}{2007}), \eprint{hep-ph/0611348}.

\bibitem[{\citenamefont{Bern et~al.}(1998)\citenamefont{Bern, Dixon, and
  Kosower}}]{Bern:1997sc}
\bibinfo{author}{\bibfnamefont{Z.}~\bibnamefont{Bern}},
  \bibinfo{author}{\bibfnamefont{L.~J.} \bibnamefont{Dixon}}, \bibnamefont{and}
  \bibinfo{author}{\bibfnamefont{D.~A.} \bibnamefont{Kosower}},
  \bibinfo{journal}{Nucl. Phys.} \textbf{\bibinfo{volume}{B513}},
  \bibinfo{pages}{3} (\bibinfo{year}{1998}), \eprint{hep-ph/9708239}.

\bibitem[{\citenamefont{Bern et~al.}(1997)\citenamefont{Bern, Dixon, Kosower,
  and Weinzierl}}]{Bern:1996ka}
\bibinfo{author}{\bibfnamefont{Z.}~\bibnamefont{Bern}},
  \bibinfo{author}{\bibfnamefont{L.~J.} \bibnamefont{Dixon}},
  \bibinfo{author}{\bibfnamefont{D.~A.} \bibnamefont{Kosower}},
  \bibnamefont{and}
  \bibinfo{author}{\bibfnamefont{S.}~\bibnamefont{Weinzierl}},
  \bibinfo{journal}{Nucl. Phys.} \textbf{\bibinfo{volume}{B489}},
  \bibinfo{pages}{3} (\bibinfo{year}{1997}), \eprint{hep-ph/9610370}.

\bibitem[{\citenamefont{Ellis and Veseli}(1999)}]{Ellis:1998fv}
\bibinfo{author}{\bibfnamefont{R.~K.} \bibnamefont{Ellis}} \bibnamefont{and}
  \bibinfo{author}{\bibfnamefont{S.}~\bibnamefont{Veseli}},
  \bibinfo{journal}{Phys. Rev.} \textbf{\bibinfo{volume}{D60}},
  \bibinfo{pages}{011501} (\bibinfo{year}{1999}), \eprint{hep-ph/9810489}.

\bibitem[{\citenamefont{Campbell and Ellis}(2000)}]{Campbell:2000bg}
\bibinfo{author}{\bibfnamefont{J.~M.} \bibnamefont{Campbell}} \bibnamefont{and}
  \bibinfo{author}{\bibfnamefont{R.~K.} \bibnamefont{Ellis}},
  \bibinfo{journal}{Phys. Rev.} \textbf{\bibinfo{volume}{D62}},
  \bibinfo{pages}{114012} (\bibinfo{year}{2000}), \eprint{hep-ph/0006304}.

\bibitem[{\citenamefont{Campbell and Ellis}(2002)}]{Campbell:2002tg}
\bibinfo{author}{\bibfnamefont{J.}~\bibnamefont{Campbell}} \bibnamefont{and}
  \bibinfo{author}{\bibfnamefont{R.~K.} \bibnamefont{Ellis}},
  \bibinfo{journal}{Phys. Rev.} \textbf{\bibinfo{volume}{D65}},
  \bibinfo{pages}{113007} (\bibinfo{year}{2002}), \eprint{hep-ph/0202176}.

\bibitem[{\citenamefont{Campbell et~al.}(2003)\citenamefont{Campbell, Ellis,
  and Rainwater}}]{Campbell:2003hd}
\bibinfo{author}{\bibfnamefont{J.}~\bibnamefont{Campbell}},
  \bibinfo{author}{\bibfnamefont{R.~K.} \bibnamefont{Ellis}}, \bibnamefont{and}
  \bibinfo{author}{\bibfnamefont{D.~L.} \bibnamefont{Rainwater}},
  \bibinfo{journal}{Phys. Rev.} \textbf{\bibinfo{volume}{D68}},
  \bibinfo{pages}{094021} (\bibinfo{year}{2003}), \eprint{hep-ph/0308195}.

\bibitem[{\citenamefont{Febres~Cordero
  et~al.}(2006)\citenamefont{Febres~Cordero, Reina, and
  Wackeroth}}]{FebresCordero:2006sj}
\bibinfo{author}{\bibfnamefont{F.}~\bibnamefont{Febres~Cordero}},
  \bibinfo{author}{\bibfnamefont{L.}~\bibnamefont{Reina}}, \bibnamefont{and}
  \bibinfo{author}{\bibfnamefont{D.}~\bibnamefont{Wackeroth}},
  \bibinfo{journal}{Phys. Rev.} \textbf{\bibinfo{volume}{D74}},
  \bibinfo{pages}{034007} (\bibinfo{year}{2006}), \eprint{hep-ph/0606102}.

\bibitem[{\citenamefont{Febres~Cordero
  et~al.}(2008)\citenamefont{Febres~Cordero, Reina, and
  Wackeroth}}]{FebresCordero:2008ci}
\bibinfo{author}{\bibfnamefont{F.}~\bibnamefont{Febres~Cordero}},
  \bibinfo{author}{\bibfnamefont{L.}~\bibnamefont{Reina}}, \bibnamefont{and}
  \bibinfo{author}{\bibfnamefont{D.}~\bibnamefont{Wackeroth}},
  \bibinfo{journal}{Phys. Rev.} \textbf{\bibinfo{volume}{D78}},
  \bibinfo{pages}{074014} (\bibinfo{year}{2008}), \eprint{arXiv:0806.0808
  [hep-ph]}.

\bibitem[{\citenamefont{Febres~Cordero}(2008)}]{Cordero:2008ce}
\bibinfo{author}{\bibfnamefont{F.}~\bibnamefont{Febres~Cordero}}
  (\bibinfo{year}{2008}), \eprint{arXiv:0809.3829 [hep-ph]}.

\bibitem[{\citenamefont{Campbell et~al.}(2004{\natexlab{b}})}]{Campbell:2004pu}
\bibinfo{author}{\bibfnamefont{J.~M.} \bibnamefont{Campbell}}
  \bibnamefont{et~al.} (\bibinfo{year}{2004}{\natexlab{b}}),
  \eprint{hep-ph/0405302}.

\bibitem[{\citenamefont{Assamagan et~al.}(2004)}]{Assamagan:2004mu}
\bibinfo{author}{\bibfnamefont{K.~A.} \bibnamefont{Assamagan}}
  \bibnamefont{et~al.} (\bibinfo{collaboration}{Higgs Working Group})
  (\bibinfo{year}{2004}), \eprint{hep-ph/0406152}.

\bibitem[{\citenamefont{Kr{\"a}mer}(2004)}]{Kramer:2004ie}
\bibinfo{author}{\bibfnamefont{M.}~\bibnamefont{Kr{\"a}mer}},
  \bibinfo{journal}{Nucl. Phys. Proc. Suppl.} \textbf{\bibinfo{volume}{135}},
  \bibinfo{pages}{66} (\bibinfo{year}{2004}), \eprint{hep-ph/0407080}.

\bibitem[{\citenamefont{Campbell
  et~al.}(2009{\natexlab{a}})\citenamefont{Campbell, Frederix, Maltoni, and
  Tramontano}}]{Campbell:2009ss}
\bibinfo{author}{\bibfnamefont{J.~M.} \bibnamefont{Campbell}},
  \bibinfo{author}{\bibfnamefont{R.}~\bibnamefont{Frederix}},
  \bibinfo{author}{\bibfnamefont{F.}~\bibnamefont{Maltoni}}, \bibnamefont{and}
  \bibinfo{author}{\bibfnamefont{F.}~\bibnamefont{Tramontano}}
  (\bibinfo{year}{2009}{\natexlab{a}}), \eprint{arXiv:0903.0005 [hep-ph]}.

\bibitem[{\citenamefont{Campbell
  et~al.}(2009{\natexlab{b}})\citenamefont{Campbell, Ellis, Febres~Cordero,
  Maltoni, Reina, Wackeroth, and Willenbrock}}]{Campbell:2008hh}
\bibinfo{author}{\bibfnamefont{J.}~\bibnamefont{Campbell}},
  \bibinfo{author}{\bibfnamefont{R.~K.} \bibnamefont{Ellis}},
  \bibinfo{author}{\bibfnamefont{F.}~\bibnamefont{Febres~Cordero}},
  \bibinfo{author}{\bibfnamefont{F.}~\bibnamefont{Maltoni}},
  \bibinfo{author}{\bibfnamefont{L.}~\bibnamefont{Reina}},
  \bibinfo{author}{\bibfnamefont{D.}~\bibnamefont{Wackeroth}},
  \bibnamefont{and}
  \bibinfo{author}{\bibfnamefont{S.}~\bibnamefont{Willenbrock}},
  \bibinfo{journal}{Phys. Rev.} \textbf{\bibinfo{volume}{D79}},
  \bibinfo{pages}{034023} (\bibinfo{year}{2009}{\natexlab{b}}),
  \eprint{arXiv:0809.3003 [hep-ph]}.

\bibitem[{\citenamefont{Campbell
  et~al.}(2009{\natexlab{c}})\citenamefont{Campbell, Ellis, Febres~Cordero,
  Maltoni, Reina, Wackeroth, and Willenbrock}}]{Zb_inprep}
\bibinfo{author}{\bibfnamefont{J.}~\bibnamefont{Campbell}},
  \bibinfo{author}{\bibfnamefont{R.~K.} \bibnamefont{Ellis}},
  \bibinfo{author}{\bibfnamefont{F.}~\bibnamefont{Febres~Cordero}},
  \bibinfo{author}{\bibfnamefont{F.}~\bibnamefont{Maltoni}},
  \bibinfo{author}{\bibfnamefont{L.}~\bibnamefont{Reina}},
  \bibinfo{author}{\bibfnamefont{D.}~\bibnamefont{Wackeroth}},
  \bibnamefont{and}
  \bibinfo{author}{\bibfnamefont{S.}~\bibnamefont{Willenbrock}}
  (\bibinfo{year}{2009}{\natexlab{c}}), \bibinfo{note}{in preparation.}

\bibitem[{\citenamefont{Campbell and Ellis}()}]{MCFM:2004}
\bibinfo{author}{\bibfnamefont{J.}~\bibnamefont{Campbell}} \bibnamefont{and}
  \bibinfo{author}{\bibfnamefont{R.~K.} \bibnamefont{Ellis}},
  \bibinfo{note}{webpage: mcfm.fnal.gov}.

\bibitem[{\citenamefont{Lai et~al.}(2000)}]{Lai:1999wy}
\bibinfo{author}{\bibfnamefont{H.~L.} \bibnamefont{Lai}} \bibnamefont{et~al.}
  (\bibinfo{collaboration}{CTEQ}), \bibinfo{journal}{Eur. Phys. J.}
  \textbf{\bibinfo{volume}{C12}}, \bibinfo{pages}{375} (\bibinfo{year}{2000}),
  \eprint{hep-ph/9903282}.

\bibitem[{\citenamefont{Catani et~al.}(1992)\citenamefont{Catani, Dokshitzer,
  and Webber}}]{Catani:1992zp}
\bibinfo{author}{\bibfnamefont{S.}~\bibnamefont{Catani}},
  \bibinfo{author}{\bibfnamefont{Y.~L.} \bibnamefont{Dokshitzer}},
  \bibnamefont{and} \bibinfo{author}{\bibfnamefont{B.~R.}
  \bibnamefont{Webber}}, \bibinfo{journal}{Phys. Lett.}
  \textbf{\bibinfo{volume}{B285}}, \bibinfo{pages}{291} (\bibinfo{year}{1992}).

\bibitem[{\citenamefont{Catani et~al.}(1993)\citenamefont{Catani, Dokshitzer,
  Seymour, and Webber}}]{Catani:1993hr}
\bibinfo{author}{\bibfnamefont{S.}~\bibnamefont{Catani}},
  \bibinfo{author}{\bibfnamefont{Y.~L.} \bibnamefont{Dokshitzer}},
  \bibinfo{author}{\bibfnamefont{M.~H.} \bibnamefont{Seymour}},
  \bibnamefont{and} \bibinfo{author}{\bibfnamefont{B.~R.}
  \bibnamefont{Webber}}, \bibinfo{journal}{Nucl. Phys.}
  \textbf{\bibinfo{volume}{B406}}, \bibinfo{pages}{187} (\bibinfo{year}{1993}).

\bibitem[{\citenamefont{Ellis and Soper}(1993)}]{Ellis:1993tq}
\bibinfo{author}{\bibfnamefont{S.~D.} \bibnamefont{Ellis}} \bibnamefont{and}
  \bibinfo{author}{\bibfnamefont{D.~E.} \bibnamefont{Soper}},
  \bibinfo{journal}{Phys. Rev.} \textbf{\bibinfo{volume}{D48}},
  \bibinfo{pages}{3160} (\bibinfo{year}{1993}), \eprint{hep-ph/9305266}.

\bibitem[{\citenamefont{Kilgore and Giele}(1997)}]{Kilgore:1996sq}
\bibinfo{author}{\bibfnamefont{W.~B.} \bibnamefont{Kilgore}} \bibnamefont{and}
  \bibinfo{author}{\bibfnamefont{W.~T.} \bibnamefont{Giele}},
  \bibinfo{journal}{Phys. Rev.} \textbf{\bibinfo{volume}{D55}},
  \bibinfo{pages}{7183} (\bibinfo{year}{1997}), \eprint{hep-ph/9610433}.

\end{thebibliography}

\end{document}